\documentclass[epj]{svjour}

\usepackage{umlaut}
\usepackage{graphics}
\usepackage{bm}
\usepackage{bbm}
\usepackage{epic}
\usepackage{eepic}
\usepackage{nicefrac}
\usepackage{amsmath}

\begin{document}

\title{Large-amplitude behavior of the Grinfeld instability: a
  variational approach}
\author{Peter Kohlert\inst{1} \and Klaus Kassner\inst{1} \and Chaouqi
  Misbah\inst{2} }

\mail{klaus.kassner@physik.uni-magdeburg.de} \offprints{Klaus Kassner}

\institute{Institut f\"ur Theoretische Physik,
  Otto-von-Guericke-Universit\"at Magdeburg, Postfach 4120, D-39016
  Magdeburg, Germany \and Groupe de Recherche sur les
  Ph{\'e}nom{\`e}nes hors de l'Equilibre, LSP, Universit{\'e} Joseph
  Fourier (CNRS), Grenoble I, B.P.\,87, Saint-Martin d'H{\`e}res,
  38402 Cedex, France}
\date{Received: date / Revised version: date}

\abstract{ In previous work, we have performed amplitude expansions of
  the continuum equations for the Grinfeld instability and carried
  them to high orders.  Nevertheless, the approach turned out to be
  restricted to relatively small amplitudes.  In this article, we use
  a variational approach in terms of multi-cycloid curves instead.
  Besides its higher precision at given order, the method has the
  advantages of giving a transparent physical meaning to the
  appearance of cusp singularities and of not being restricted to
  interfaces representable as single-valued functions.  Using a single
  cycloid as ansatz function, the entire calculation can be performed
  analytically, which gives a good qualitative overview of the system.
  Taking into account several but few cycloid modes, we obtain
  remarkably good quantitative agreement with previous numerical
  calculations.  With a few more modes taken into consideration, we
  improve on the accuracy of those calculations. Our approach extends
  them to situations involving gravity effects.  Results on the shape
  of steady-state solutions are presented at both large stresses and
  amplitudes.  In addition, their stability is investigated.
   }

\maketitle

\section{\label{sec:introduction}Introduction}

When a nonhydrostatically strained solid has a surface, at which
material can be redistributed by some appropriate transport mechanism,
it may reduce its elastic energy via surface undulations. Intuitively,
this should be clear: stresses are partially relieved in the maxima of
corrugations and enhanced in their minima.  The elastic energy density
is therefore reduced in the maxima and increased in the minima,
favoring growth of the former and deepening of the latter. This
mechanism is at the origin of a morphological instability leading to
the formation of grooves with a relatively well-defined initial
spacing under uniaxial stress \cite{grinfeld:86,grinfeld:93} and,
possibly, the evolution of islands, if the stress is biaxial
\cite{grinfeld:94,berger:03}.  Pertinent transport processes are
melting-crystallization for a solid in contact with its melt and
surface diffusion for a sufficiently hot solid in vacuum.  The
latter case is relevant in epitaxial growth, where the lattice
mismatch between different materials is the source of biaxial stress.

The instability seems to first have been predicted by Asaro and Tiller
\cite{asaro:72}, but its universal nature was recognized by Grinfeld
\cite{grinfeld:86}, hence it has often been referred to as Grinfeld or
Asaro-Tiller-Grinfeld (ATG) instability.  An unambiguous experimental
demonstration of the instability was given by Torii and Balibar
\cite{torii:92}, using solid helium in contact with its superfluid.

It should be emphasized that the surface undulation evolving as a
consequence of the instability is not due to elastic deformation such
as bending (as would be the case on application of a pressure to a
long thin rod, leading to the Euler buckling instability). Instead, the
instability materializes itself via mass transport and is independent
of whether the stress is tensile or compressive.  When the solid is in
contact with its melt, the latter is a particle reservoir, rendering
mass transport easy (and the dynamics is not governed by a
conservation law).  When the solid is in contact with vacuum such as
in the case of heteroepitaxy, the instability takes place via surface
diffusion in most cases, but may also be supported by other transport
processes such as vacancy or impurity diffusion.  In that case, mass
conservation is important in the dynamics. For a pedagogical
introduction into the subject of the ATG instability, we refer the
reader to \cite{cantat:98}.

There are a number of interesting questions concerning the instability.
Since it produces crack-like patterns \cite{yang:93}, does it
constitute a generic route to fracture as hypothesized in
\cite{kassner:94} or will plasticity in general lead to a
restabilization? If one restricts oneself to linear elasticity, are
there steady states beyond those found by Spencer and Meiron
\cite{spencer:94}? It has been shown that in directional solidification
stable steady state patterns are realizable \cite{cantat:98,kappey:00}.
For the pure Grinfeld instability, this appears to be impossible in
extended systems. Further questions concern the nature of dynamical
states, when there is no steady state. Coarsening has been found to be
a generic behaviour \cite{mueller:98,mueller:99,kassner:01}, but more
detailed investigations on large-scale systems would be desirable, to
determine the precise form of the pertinent power laws.

Numerical simulations of a solid undergoing the Grinfeld instability
\cite{yang:93,kassner:94,spencer:94} have the awkward tendency of
producing cusp singularities in finite time.  The investigation of
Spencer and Meiron \cite{spencer:94} has shown that these
singularities are not an artifact of the numerics but intrinsic to the
continuum model describing the system, under the assumption of linear
elasticity (and in the limit of negligible sound propagation effects).
What they found was a steady branch of solutions in a certain range of
wavelengths, corresponding to very small sinusoidal shapes near the
onset of the branch and approaching a cycloid-like cuspy shape near
its termination.

Such a result might have been anticipated on the basis of the analytic
work by Chiu and Gao \cite{gao:93}, who performed a detailed
calculation of the stress state under a cycloid-shaped surface using
the Goursat function scheme proposed by Mu{\ss}chelischwili
\cite{mus:1}. All of these numerical studies considered
two-dimensional systems where the interface is described by a curve.
Whereas we have treated three-dimensional systems in \cite{berger:03}
within a weakly nonlinear approach, we will restrict ourselves to two
dimensions here but go well beyond the regime of validity of weakly
nonlinear theory.

Chiu and Gao find that for a certain range of wavenumbers a fully
cusped cycloid constitutes an energetically more stable configuration
than a flat surface.  In section \ref{sec:monocyc}, we will show that
a variational calculation using cycloids as ansatz functions gives a
rather good approximation of steady state solutions of arbitrary
amplitude, some of which were discussed in \cite{spencer:94}.

Moreover, we are able to draw conclusions on the large amplitude
behavior even for strong gravity or a large density difference between
the solid and nonsolid phases (liquid or vacuum), as we show in
section \ref{sec:gravity}. Evidence for these states has already been
found in \cite{kohlert:02a}.

In section \ref{sec:multicyc}, we present a generalization of this
idea.  Employing a special system of (not necessarily
univalent) functions called multi-cycloids we analytically recover the
numerical results for the mean square amplitude to excellent accuracy
already at third order. At higher order, we get more precise results
with less numerical effort than Ref.~\cite{spencer:94}.

Finally, we give some conclusions as to the physical interpretation of
our results and suggest how to verify them experimentally or by a full
numerical computation.

\section{\label{sec:monocyc}The mono-cycloid approximation}
\subsection{Cycloids}
We wish to use cycloids and more general curves deriving from cycloids
to model the steady-state surface pattern of a two-dimensional solid
after it has undergone the Grinfeld instability. This is of course
motivated by the fact that cycloids have been shown by Chiu and Gao
\cite{gao:93} to be very efficient in reducing the elastic energy, and
hence the final steady state, if any, should be close to a cycloid
shape.
%The same motivation led Spencer and Meiron to using
%cycloid-like functions in parametrizing the discretization points
%within their fully numerical approach \cite{spencer:94}.

Cycloids belong to the more generic class of trochoids, curves defined
as the trace of a point fixed on a circle rolling along some
prescribed line.  A cycloid is the curve traced out by a point on the
circumference of a circle as the latter rolls along a straight line.
When we put the point inside or outside the circle instead, we obtain
a curtate or a prolate cycloid, respectively. The parametric
representation of a cycloid can be given in a compact manner by a
complex generating function
\begin{equation}
\label{eq:monogenerator} \zeta(\xi)=\xi-i\frac{\rho}{k} e^{-ik\xi}\>,
\end{equation}
Herein, $i$ is the imaginary unit, $k$ is a wavenumber and $\rho$ is a
dimensionless amplitude-like parameter. The cycloid is obtained by
taking the real and imaginary part as the $x$ and $y$ coordinates,
respectively.

\begin{figure}[htb]
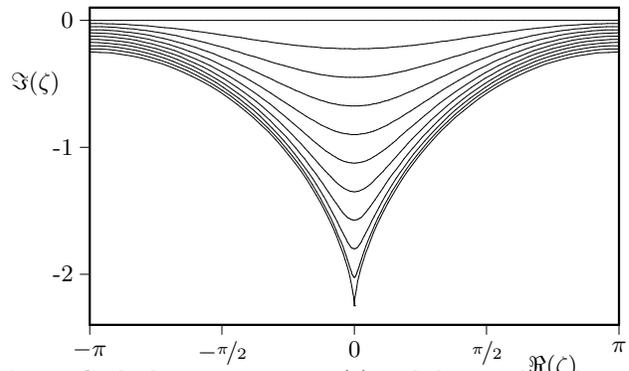

\setlength{\unitlength}{\columnwidth}

\caption{\label{fig:zykloform}Cycloids, as in equation
  \eqref{eq:monogenerator} with $k=1$, plotted in the range
  $\xi=(-\pi,\pi)$; $\rho=0\dots 1$ from top to bottom. Curves have
  been shifted in order to avoid overlapping.}
\end{figure}
Taking $\rho=1$ leads to the representation of the classical, i.e.,
cusped, cycloid.  Choice of a plus instead of the minus sign in
equation \eqref{eq:monogenerator} would just shift the minimum from
$\xi=0$ to $\xi=\pi/k$, while a plus sign in the exponent would lead
to a surface with the cusps pointing upward. The latter case is of no
relevance for the Grinfeld instability, because a fully relaxed upward
cusp would immediately shrink under any perturbation in order to
decrease the surface energy.

For $\rho>1$, the cycloid becomes self-intersecting and does not
represent a physical state anymore. Note, however, that if we
``superimpose'' several cycloids\footnote{This is not a superposition
  in the standard sense. We add up partial representations of the $x$ and
  $y$ coordinates, so the whole curve is not a simple sum, see
  Eq.~\eqref{eq:multigenerator}.}
as we will do in section \ref{sec:multicyc}, the possibility of the $x$
coordinate to vary nonmonotonously allows the representation of
patterns with overhangs that do not self-intersect. We will not discuss
this feature in detail here but report on those patterns in a different
article.

The parametric representation of the cycloid is
\begin{subequations}
\label{eq:xy}
\begin{align}
\label{eq:x} x(\xi)= &\Re[\zeta(\xi)]=
\xi-\frac{\rho}{k}\sin{\left(k\xi\right)}\>,\\
\label{eq:y} y(\xi)= &\Im[\zeta(\xi)]=-\frac{\rho}{k}\cos{\left(k\xi\right)}
\end{align}
\end{subequations}
Assuming the surface of a solid having undergone the Grinfeld
instability to be described by this shape, we shift the cycloid
position by its mean value
\begin{equation}
m=\frac{k}{2\pi}\int\dot{x}(\xi)y(\xi){\rm d}\xi=\frac{\rho^2}{2k}\>,
\end{equation}
in order to keep the average interface position at $y=0$. Herein, the
dot denotes differentiation with respect to $\xi$. Hence, from now on we
use
\begin{equation}
\label{eq:y_gesenkt} y(\xi)=-\frac{\rho}{k}\cos{\left(k\xi\right)}
 - \frac{\rho^2}{2k}\>.
\end{equation}
Moreover, if we want to compare results with both our amplitude
calculation \cite{kohlert:02a} and the work of Spencer and Meiron
\cite{spencer:94}, we additionally have to know the mean square
amplitude of the cycloid. As in \cite{kohlert:02a}, we denote this
mean square amplitude by $\overline{\alpha}\>$; its calculation yields
\begin{equation}
\label{eq:quadratmittel}
\overline{\alpha}=\left[\frac{k}{2\pi}\int\dot{x}(\xi)y(\xi)^2{\rm
d}\xi\right]^{\nicefrac{1}{2}}=\left| \frac{\rho\sqrt{2-\rho^2}}{2k}\right|\>,
\end{equation}
where the absolute value ensures a nonnegative mean square amplitude,
no matter what the sign of $\rho$ and of $k$.

\subsection{Scaling}
Our basic model is an isotropic solid obeying the laws of linear
elasticity (i.e., the Lamé equations) with a surface on which shear
stresses vanish while the normal stress component is equal to the
negative pressure in the liquid or zero. That is, we neglect capillary
overpressure due to a curved interface, which is known to be a small
cross effect \cite{cantat:98}. Moreover, we neglect the body force
effect of gravity in the elastic equations, also known to be small.

The energy of the solid then consists of three contributions, its
elastic energy, total surface energy, and potential energy in the
gravity field.

It is known \cite{kassner:01} that if the latter force, gravity, is
completely neglected, the equations of motion of the Grinfeld
instability can be made parameter free. This is achieved by referring
all lengths to a length scale $l_1$, essentially the Griffith length,
given by:
\begin{equation}
l_1=\frac{\gamma}{2w_0} \>,
\end{equation}
where $\gamma$ is the surface tension and $w_0=\sigma_0^2(1-\nu^2)/2E$
the elastic energy density of the prestressed planar state.
$\sigma_0=\sigma_{xx}-\sigma_{zz}$ is the first normal stress difference
or, in more physical terms, the excess stress applied in the $x$
direction, to produce a uniaxially strained solid, whereas $E$ and
$\nu$ are elastic constants describing an elastically isotropic
material, {\em viz.} Young's modulus and the Poisson number.

Physically, the Griffith length describes the competition between
surface energy and elastic energy. It is used predominantly in the
theory of crack propagation. Cracks larger than this length relieve
more elastic energy when growing than they produce surface energy,
while cracks shorter than it can reduce energy only by shrinking.
Therefore, this length scale represents a nucleation size for crack
generation.

When gravity is considered, another length $l_2$ becomes important,
which is
\begin{equation}
l_2=\frac{w_0}{g \Delta\rho}\>.
\end{equation}
Herein, $\Delta\rho$ is the density difference between the solid and
the second phase, $g$ the gravitational acceleration. This gravity
length describes competition between elastic energy and potential
energy in a gravitational field. Due to the density difference between
the two phases, the system can gain potential energy, if the phase
with the larger density, usually the solid, shifts its center of mass
downward. As a consequence, if a solid immersed in and in equilibrium
with its melt is submitted to a uniaxial stress, it will first start
to melt, because it is now out of equilibrium; but then its center of
mass shifts downward, hence the potential energy is decreased and a
new equilibrium state may be reached.  This happens whenever the
applied stress difference is below the instability threshold. The
solid surface melts back by a certain height, and this height change
is exactly given by $l_2$.

 The only parameter of the nondimensionalized equations is
the ratio of these length scales:
\begin{equation}
l_{12}:=\frac{l_1}{2\,l_2}\>.
\end{equation}
In all considerations that follow, we have carried out a formal
transformation $x\rightarrow x/l_1$, $y\rightarrow y/l_1$,
$k\rightarrow kl_1$, $\xi\rightarrow\xi/l_1$, and energies and their
variations have been divided by a common prefactor $\gamma$.

\subsection{The cycloid approximation for the no-gravity case}
In treating the cycloid approximation, we choose an approach that is
essentially variational in nature.  Therefore, we need not compute the
energy itself but only its variation.  The variation of strain energy
due to a configurational variation $\mathbf{\delta x}$ may be written
as
\begin{equation}
\delta E_{\rm e}=\int w(s)\mathbf{n}\,\mathbf{\delta x}\,{\mathrm d}s. \label{eq:du}
\end{equation}
where $w(s)$ is the energy density at the surface, $\mathbf{n}$ is the
normal vector and $s$ denotes the arclength along the surface (as we
are dealing with a two-dimensional system). Using Eqs.~\eqref{eq:x} and
\eqref{eq:y_gesenkt}, we can calculate an approximation to $\delta
E_{\rm e}$, allowing only the parameter $\rho$ to vary (instead of
taking the full variational derivative which would take the result out
of the space of cycloidal shapes)
\begin{equation}
\begin{split}
\mathbf{\delta x}=&\left(\frac{\partial x(\xi)}{\partial
\rho}\mathbf{e}_x +\frac{\partial y(\xi)}{\partial
\rho}\mathbf{e}_y\right)\delta\rho\\=&-k^{-1}\left[\sin(k\xi)\mathbf{e}_x
+\left(\cos(k\xi)+\rho\right)\mathbf{e}_y\right]\delta\rho\>,
\label{eq:dx}
\end{split}
\end{equation}
and we have the relation
\begin{equation}
\begin{split}
\mathbf{n}ds=&\left(-\dot{y}\left(\xi\right) \mathbf{e}_x+\dot{x}\left(\xi\right)
\mathbf{e}_y\right)d\xi\\
=&\left[\left(-\rho\sin(k\xi)\right)\mathbf{e}_x
+\left(1-\rho\cos(k\xi)\right)\mathbf{e}_y\right]d\xi \>.
\end{split}
\label{eq:nds}
\end{equation}
The calculation of the strain energy density can be performed for the
general case of a multi-cycloid. Essentially the same calculation has
been carried out before by Yu and Suo \cite{yu:99}, who used it to
model groove-to-crack evolution in ceramics, a context quite different
from ours.  Since the notations used by the two groups are pretty
different and a direct translation would be tedious, we present the
important steps of our calculation (done independently and based on
\cite{kassner:01} rather than \cite{yu:99}) in appendix
\ref{app:generalmapping}. At this point we only need the energy
density for the mono-cycloidal surface:
\begin{equation}
w(s)=\frac{1}{2} \frac{ \left( 1-\rho^2 \right) ^2}{ \left(1+\rho^2 -2\rho\cos \left( k
\xi \right)  \right)^2} \>.
\end{equation}
When comparing with the results of \cite{gao:93}, one finds the only
difference (up to prefactors) in the different sign of the cosine
function, which results from their different ansatz of the generating
function, corresponding to a simple shift of the argument of the
cosine.  The integration yields a surprisingly simple result:
\begin{equation}
\label{eq:dEe}\frac{\partial E_{\rm e}}{\partial\rho}=\frac{1}{k}\int w(s) \cos \left(
k\xi \right)  \left(\rho^2-1 \right)  {\rm d}\xi =-\frac{2\pi\rho}{k^2}\>,
\end{equation}
where we have taken $k>0$ (otherwise the expression on the right-hand
side would have to be multiplied by the sign of $k$).

It remains to be noticed that the integration can be done analytically in the
mono-cycloid case, yielding
\begin{equation}
\label{eq:integrated} E_{\rm e}=-\frac{\pi\rho^2}{k^2}\>.
\end{equation}

The surface energy in our scalings simply is the difference of the arc
lengths, and its derivative is calculated straightforwardly:
\begin{subequations}
\begin{equation}\label{eq:bogenlaenge}
E_{\rm s}=\int\sqrt{\dot{x}\left(\xi\right)^2+\dot{y}\left(\xi\right)^2}
{\rm d}\xi-1 \>,
\end{equation}
\begin{equation}
E_{\rm s}=\frac{4\left(1+\rho\right)}{k} {\rm E} \left({\frac {2\sqrt {
\rho}}{1+\rho}}\right)-1      \>, \end{equation}
\end{subequations}
\begin{equation} \label{eq:dEs}
\frac{\partial E_{\rm s}}{\partial\rho}=\frac{2}{k}\left[ \frac{\rho-1}{\rho}{\rm
K}\left( \frac { 2\sqrt {\rho}}{1+\rho} \right)+\frac{\rho+1}{\rho}{\rm
E}\left(\frac{2\sqrt{\rho}}{1+\rho} \right)\right] \>.
\end{equation}
Herein, $\rm K$ and $\rm E$ are complete elliptic integrals of the
first and second kind, respectively, defined by
\begin{subequations}
\begin{align}
K(u)&= \int_0^{\nicefrac{\pi}{2}} \frac{dx}{\sqrt{1-u^2\sin^2 x}}\>,
& (\vert u\vert &<1)\>, \\
E(u)&= \int_0^{\nicefrac{\pi}{2}} dx {\sqrt{1-u^2\sin^2 x}}\>, &
(\vert u\vert& \le 1)\>.
\end{align}
\end{subequations}
The result \eqref{eq:dEs} simplifies to
 $\nicefrac{\partial E_s}{\partial \rho} =
\nicefrac{4}{k}$ in the fully cusped limit, i.e., for $\rho=1$.
For later simplifications we set
\[N(\rho):=\frac{2}{\pi}\left(\left(\rho-1\right){\rm
K}\left({\frac{2\sqrt{\rho}}{1+\rho}}\right)+ \left(\rho+1\right){\rm
E}\left({\frac{2\sqrt{\rho}}{1+\rho}}\right)\right)\>,\] leading to
\begin{equation} \label{eq:dEs2}
\frac{\partial E_{\rm s}}{\partial\rho}=\frac{N(\rho)\pi}{k\rho}\>.
\end{equation}
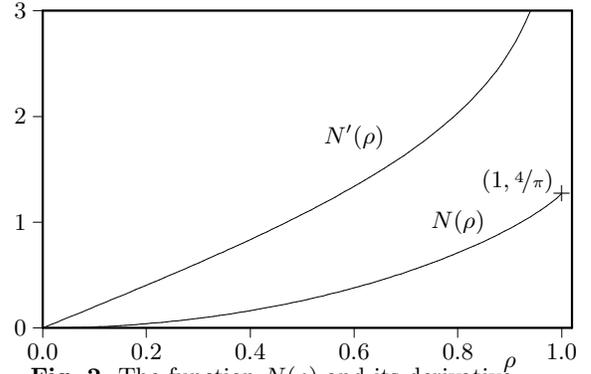
\begin{figure}[htb]
\setlength{\unitlength}{\columnwidth}
\begin{picture}(1.000000,.510000)(-.150000,-.030000)
\allinethickness{.040000mm} \thinlines \thicklines
\drawline(0,0)(0,.480000)(.800000,.480000)(.800000,0)(0,0) \thinlines
\drawline(0.000000,0.000000)(-.015000,0.000000)
\put(-.025000,0.000000){\makebox(0,0)[r]{0}}
\drawline(0.000000,.160000)(-.015000,.160000)
\put(-.025000,.160000){\makebox(0,0)[r]{1}}
\drawline(0.000000,.320000)(-.015000,.320000)
\put(-.025000,.320000){\makebox(0,0)[r]{2}}
\drawline(0.000000,.480000)(-.015000,.480000)
\put(-.025000,.480000){\makebox(0,0)[r]{3}}
\put(-.045000,.400000){\makebox(0,0)[r]{$$}}
\drawline(0.000000,0.000000)(0.000000,-.015000)
\put(0.000000,-.025000){\makebox(0,0)[t]{0.0}}
\drawline(.156863,0.000000)(.156863,-.015000)
\put(.156863,-.025000){\makebox(0,0)[t]{0.2}}
\drawline(.313725,0.000000)(.313725,-.015000)
\put(.313725,-.025000){\makebox(0,0)[t]{0.4}}
\drawline(.470588,0.000000)(.470588,-.015000)
\put(.470588,-.025000){\makebox(0,0)[t]{0.6}}
\drawline(.627451,0.000000)(.627451,-.015000)
\put(.627451,-.025000){\makebox(0,0)[t]{0.8}}
\drawline(.784314,0.000000)(.784314,-.015000)
\put(.784314,-.025000){\makebox(0,0)[t]{1.0}}
\put(.705882,-.045000){\makebox(0,0)[t]{$\rho$}} \thinlines
\drawline(0.000000,0.000000)(.007843,.000016)(.015686,.000064)
(.023529,.000144)(.031373,.000256)(.039216,.000400)(.047059,.000576)
(.054902,.000784)(.062745,.001025)(.070588,.001297)(.078431,.001602)
(.086275,.001939)(.094118,.002308)(.101961,.002710)(.109804,.003144)
(.117647,.003610)(.125490,.004109)(.133333,.004641)(.141176,.005205)
(.149020,.005802)(.156863,.006432)(.164706,.007096)(.172549,.007792)
(.180392,.008521)(.188235,.009284)(.196078,.010080)(.203922,.010910)
(.211765,.011773)(.219608,.012671)(.227451,.013602)(.235294,.014568)
(.243137,.015568)(.250980,.016602)(.258824,.017671)(.266667,.018776)
(.274510,.019915)(.282353,.021089)(.290196,.022300)(.298039,.023545)
(.305882,.024827)(.313725,.026146)(.321569,.027500)(.329412,.028892)
(.337255,.030320)(.345098,.031786)(.352941,.033290)(.360784,.034832)
(.368627,.036412)(.376471,.038030)(.384314,.039688)(.392157,.041385)
(.400000,.043122)(.407843,.044900)(.415686,.046718)(.423529,.048577)
(.431373,.050478)(.439216,.052421)(.447059,.054407)(.454902,.056436)
(.462745,.058509)(.470588,.060627)(.478431,.062789)(.486275,.064998)
(.494118,.067253)(.501961,.069555)(.509804,.071906)(.517647,.074305)
(.525490,.076755)(.533333,.079255)(.541176,.081807)(.549020,.084412)
(.556863,.087071)(.564706,.089786)(.572549,.092558)(.580392,.095387)
(.588235,.098277)(.596078,.101227)(.603922,.104241)(.611765,.107321)
(.619608,.110467)(.627451,.113683)(.635294,.116972)(.643137,.120335)
(.650980,.123777)(.658824,.127300)(.666667,.130910)(.674510,.134609)
(.682353,.138404)(.690196,.142300)(.698039,.146304)(.705882,.150424)
(.713725,.154670)(.721569,.159054)(.729412,.163590)(.737255,.168297)
(.745098,.173200)(.752941,.178336)(.760784,.183756)(.768627,.189552)
(.776471,.195912)(.784314,.203718)
\drawline(0.000000,0.000000)(.007372,.003008)(.014745,.006016)
(.022117,.009026)(.029489,.012036)(.036862,.015048)(.044234,.018062)
(.051606,.021078)(.058979,.024097)(.066351,.027120)(.073723,.030146)
(.081096,.033176)(.088468,.036211)(.095840,.039250)(.103213,.042295)
(.110585,.045345)(.117957,.048402)(.125330,.051466)(.132702,.054536)
(.140074,.057614)(.147447,.060701)(.154819,.063795)(.162191,.066899)
(.169564,.070012)(.176936,.073136)(.184308,.076270)(.191681,.079414)
(.199053,.082571)(.206425,.085740)(.213798,.088921)(.221170,.092116)
(.228542,.095325)(.235915,.098549)(.243287,.101787)(.250659,.105042)
(.258032,.108313)(.265404,.111602)(.272777,.114908)(.280149,.118234)
(.287521,.121579)(.294894,.124944)(.302266,.128331)(.309638,.131740)
(.317011,.135171)(.324383,.138627)(.331755,.142108)(.339128,.145614)
(.346500,.149148)(.353872,.152710)(.361245,.156302)(.368617,.159924)
(.375989,.163578)(.383362,.167265)(.390734,.170987)(.398106,.174745)
(.405479,.178541)(.412851,.182376)(.420223,.186252)(.427596,.190172)
(.434968,.194136)(.442340,.198148)(.449713,.202209)(.457085,.206321)
(.464457,.210488)(.471830,.214712)(.479202,.218995)(.486574,.223341)
(.493947,.227753)(.501319,.232235)(.508691,.236790)(.516064,.241422)
(.523436,.246136)(.530808,.250937)(.538181,.255829)(.545553,.260819)
(.552925,.265913)(.560298,.271117)(.567670,.276439)(.575042,.281886)
(.582415,.287468)(.589787,.293196)(.597159,.299079)(.604532,.305131)
(.611904,.311365)(.619276,.317798)(.626649,.324446)(.634021,.331332)
(.641393,.338477)(.648766,.345909)(.656138,.353660)(.663510,.361767)
(.670883,.370275)(.678255,.379235)(.685627,.388712)(.693000,.398786)
(.700372,.409554)(.707744,.421144)(.715117,.433718)(.722489,.447495)
(.729861,.462779)(.737234,.480000)
\put(.627451,.160000){\makebox(0,0){$N(\rho)$}}
\put(.470588,.288000){\makebox(0,0){$N'(\rho)$}}
\put(.784314,.203718){\makebox(0,0){$+$}}
\put(.784314,.203718){\makebox(0,0)[rb]{$(1,\nicefrac{4}{\pi})\,\,$}}
\end{picture}
\caption{\label{fig:nanddn}The function $N(\rho)$ and its derivative.}
\end{figure}

In Fig.~\ref{fig:nanddn}, we plot the function $N(\rho)$ and its
derivative $dN(\rho)/d\rho=4\,\rho\, {\rm K}( 2\sqrt {\rho}/(1+\rho))
/[\pi(1+\rho)]$. First, we note that $N(\rho)$ is monotonously
increasing from $N(0)=0$ to $N(1)=\nicefrac{4}{\pi}$.  Second, the
derivative diverges logarithmically near $\rho=1$. At $\rho=0$,
$N(\rho)$ is regular, and the first few terms of its Taylor series are
given by
\begin{equation}  \label{eq:tayln}
N(\rho) = \rho^2 + \frac18 \rho^4 + \frac{3}{64} \rho^6 + \frac{25}{1024} \rho^8 +
\mathcal{O}(\rho^{10}) \>.
\end{equation}
This expansion will become useful later in the discussion of the type
of bifurcation at the instability threshold.

\begin{figure}[htb]
\setlength{\unitlength}{\columnwidth}
\begin{picture}(1.000000,.510000)(-.150000,-.030000)
\allinethickness{.100000mm} \thinlines \thicklines
\put(0.000000,0.000000){\vector(1,0){.800000}}
\put(0.000000,0.000000){\vector(0,1){.480000}} \thinlines
\drawline(0.000000,0.000000)(-.015000,0.000000)
\put(-.025000,0.000000){\makebox(0,0)[r]{0.0}}
\drawline(0.000000,.141176)(-.015000,.141176)
\put(-.025000,.141176){\makebox(0,0)[r]{0.1}}
\drawline(0.000000,.282353)(-.015000,.282353)
\put(-.025000,.282353){\makebox(0,0)[r]{0.2}}
\drawline(0.000000,.423529)(-.015000,.423529)
\put(-.025000,.423529){\makebox(0,0)[r]{0.3}}
\put(-.045000,.352941){\makebox(0,0)[r]{$\overline{\alpha}$}}
\drawline(.035556,0.000000)(.035556,-.015000)
\put(.035556,-.025000){\makebox(0,0)[t]{1.6}}
\drawline(.213333,0.000000)(.213333,-.015000)
\put(.213333,-.025000){\makebox(0,0)[t]{1.7}}
\drawline(.391111,0.000000)(.391111,-.015000)
\put(.391111,-.025000){\makebox(0,0)[t]{1.8}}
\drawline(.568889,0.000000)(.568889,-.015000)
\put(.568889,-.025000){\makebox(0,0)[t]{1.9}}
\drawline(.746667,0.000000)(.746667,-.015000)
\put(.746667,-.025000){\makebox(0,0)[t]{2.0}}
\put(.657778,-.045000){\makebox(0,0)[t]{$k$}}
\drawline(.746667,0.000000)(.746613,.004991)(.746489,.009982)
(.746267,.014972)(.745955,.019961)(.745555,.024949)(.745065,.029935)
(.744486,.034918)(.743818,.039899)(.743059,.044877)(.742211,.049851)
(.741272,.054822)(.740244,.059788)(.739124,.064750)(.737912,.069707)
(.736610,.074659)(.735215,.079605)(.733728,.084545)(.732148,.089479)
(.730475,.094406)(.728708,.099325)(.726846,.104237)(.724889,.109141)
(.722837,.114037)(.720688,.118924)(.718442,.123801)(.716098,.128669)
(.713655,.133527)(.711113,.138375)(.708470,.143211)(.705727,.148037)
(.702880,.152851)(.699931,.157653)(.696877,.162442)(.693718,.167218)
(.690452,.171981)(.687078,.176731)(.683594,.181466)(.680001,.186187)
(.676295,.190892)(.672475,.195582)(.668541,.200256)(.664490,.204914)
(.660320,.209555)(.656030,.214179)(.651618,.218785)(.647082,.223373)
(.642420,.227942)(.637629,.232492)(.632708,.237023)(.627654,.241533)
(.622464,.246024)(.617135,.250493)(.611666,.254941)(.606052,.259367)
(.600291,.263771)(.594380,.268153)(.588315,.272511)(.582092,.276845)
(.575707,.281156)(.569157,.285442)(.562437,.289703)(.555542,.293939)
(.548467,.298149)(.541207,.302333)(.533758,.306491)(.526111,.310622)
(.518262,.314726)(.510203,.318802)(.501927,.322851)(.493426,.326872)
(.484692,.330865)(.475716,.334830)(.466487,.338767)(.456995,.342676)
(.447227,.346557)(.437173,.350411)(.426816,.354237)(.416142,.358037)
(.405135,.361811)(.393774,.365559)(.382040,.369285)(.369909,.372988)
(.357354,.376670)(.344346,.380336)(.330851,.383986)(.316830,.387626)
(.302237,.391259)(.287020,.394892)(.271116,.398533)(.254449,.402190)
(.236928,.405875)(.218438,.409606)(.198833,.413402)(.177919,.417292)
(.155436,.421318)(.131002,.425539)(.104031,.430053)(.073490,.435029)
(.037128,.440845)
\drawline(.284444,.443253)(.292346,.439360)(.302222,.434948)
(.312099,.430796)(.321975,.426644)(.331852,.421453)(.341728,.416782)
(.351605,.412630)(.361481,.407958)(.371358,.403287)(.381235,.399135)
(.391111,.394464)(.400988,.389273)(.410864,.384083)(.420741,.378893)
(.430617,.373702)(.440494,.368512)(.450370,.363322)(.460247,.358131)
(.470123,.352941)(.480000,.347751)(.489877,.342561)(.499753,.337370)
(.509630,.332180)(.519506,.326471)(.529383,.320242)(.539259,.313235)
(.549136,.306747)(.559012,.299481)(.568889,.293253)(.578765,.285467)
(.588642,.278201)(.598519,.269896)(.608395,.262111)(.618272,.252768)
(.628148,.243426)(.638025,.233564)(.647901,.224740)(.657778,.214360)
(.667654,.202422)(.677531,.189446)(.687407,.176471)(.697284,.162311)
(.707160,.147114)(.717037,.127391)(.726914,.104263)(.736790,.075488)
(.741728,.051903)(.745679,.023128)(.746667,0.000000)
\put(.124444,.381176){\makebox(0,0)[rb]{m}}
\put(.480000,.395294){\makebox(0,0)[rb]{s}}
\end{picture}
\caption{\label{fig:monozyk}The solution of the mono-cycloid model (m)
  in comparison with the digitized and scaled solution branch of
  Spencer and Meiron (s).}
\end{figure}
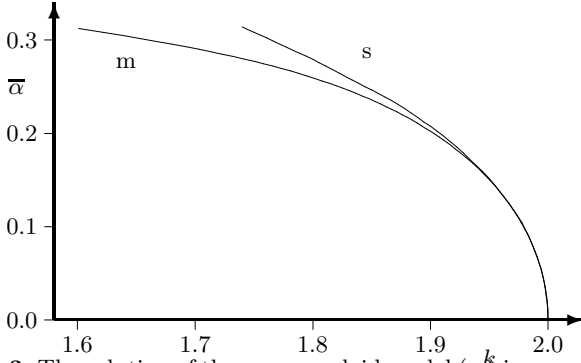

To obtain the energy minimum, we simply have to solve
\begin{equation}
\label{eq:erste} \frac{\partial E_{\rm e}}{\partial\rho}+\frac{\partial E_{\rm
s}}{\partial\rho}=\frac{2\pi}{k} \left[-\frac{\rho}{k}+\frac{N(\rho)}{2\rho}\right]=0\>.
\end{equation}
The complete branch is obtained by solving Eq.~\eqref{eq:erste} for
$k$ instead of $\rho$, taking into account that $\rho$ is running from
zero to one:
\begin{equation}
\label{eq:kmono} k=\frac{2\rho^2}{N(\rho)} \>.
\end{equation}
The solution, additionally converted to $\overline{\alpha}$ via
equation \eqref{eq:quadratmittel} for easier comparison, is shown in
figure \ref{fig:monozyk}. At the termination point of the numerical
solution \cite{spencer:94}, the $\overline{\alpha}$ value of the
mono-cycloid approximation is about 10\% smaller.

{From} \eqref{eq:kmono}, we conclude that the analytical termination
point, given by the cusp limit $\rho\!=\!1$, is located at
$k\!=\!\nicefrac{\pi}{2}$, which is somewhat different from the
termination point of Spencer and Meiron \cite{spencer:94} at about
$1.74$ (their scalings imply double wave number and half amplitude
with respect to ours, so they gave the termination point at $k=3.48$).
We shall see later that the wavenumber of the true cusp is closer to
the monocycloid result than to the numeric value. This is due to the
fact that a very small change of tip radius in the minimum of the
pattern leads to a relatively large change of the wave number. So the
Spencer-Meiron result is accurate for the part of the solution branch
that it reproduces, but it misses a considerable piece of the branch.

{From} Eq.~\eqref{eq:quadratmittel} we find using $k_{\rm
  term}=\nicefrac{\pi}{2}$
\begin{equation}
\overline{\alpha}_{\rm term}=\frac{1}{\pi}\>.
\end{equation}

An important question is the stability of our solutions. Again, only a
simple calculation needs to be performed: Take $k$ from equation
\eqref{eq:kmono}, and insert it into the second derivative of the
total energy $E=E_{\rm e}+E_{\rm s}$ . This leads to
\begin{multline}
\frac{k}{2\pi} \frac{\partial^2E}{\partial\rho^2}
=-\frac{1}{k}-\frac{N(\rho)}{2\rho^2}+\frac{N'(\rho)}{2\rho}
=-\frac{N(\rho)}{\rho^2}+\frac{N'(\rho)}{2\rho} \\
=\frac{2}{\pi}\left[\frac{1+\rho}{\rho^2}
{\rm E}\left(\frac{2\sqrt{\rho}}{1+\rho}\right)
+\frac{1}{(1+\rho)\rho^2}
{\rm K}\left(\frac{2\sqrt{\rho}}{1+\rho}\right)\right]\>,
\end{multline}
which is positive for any $\rho\in\left(0,1\right)$. Hence the
complete solution branch is stable up to the singularity, a result
that agrees with that of Spencer and Meiron \cite{spencer:94}. Of
course, with our method statements about stability can be made only
concerning the restricted set of functions used in the variational
ansatz (which depends on just one parameter here).

Beyond $\rho=1$, Eq.~\eqref{eq:erste} still gives us a stationary
point of the energy (even a minimum for $\rho$ not too far above 1),
but the corresponding cycloid self-intersects, hence the solution is
unphysical. Therefore, the variational ansatz provides a transparent
analytic explanation of the fact that the solution branch terminates.

\section{\label{sec:gravity}Including gravity}
The next natural step is the incorporation of gravity into the model.
Again, the calculation of the corresponding energy contribution is
fairly easy, because up to a prefactor it is nothing but the square of
the mean square amplitude.
\begin{equation}
E_{\rm g}=\frac{l_{12}}{2}\int\dot{x}(\xi)y(\xi)^2{\rm d}\xi =
\frac{\pi l_{12}}{4}\frac
{\rho^2\left(2-\rho^2\right)}{k^3}
\end{equation}
Consequently, the derivative is
\begin{equation}
\frac{\partial E_{\rm g}}{\partial\rho}= \frac{\pi
l_{12}\rho\left(1-\rho^2\right)}{k^3}\>,
\end{equation}
and for the generalized model we replace Eq.~\eqref{eq:erste} by
\begin{equation}
\label{eq:zweite}\frac{\partial}{\partial\rho}\left(E_{\rm e}+E_{\rm s}+E_{\rm
g}\right)=0\>.
\end{equation}
Again it turns out that the solution can be written down exactly if we
express $k$ through $\rho$.  Now we have two solutions which are both
meaningful:
\begin{equation}
\label{eq:k_extended} k=\frac{\rho^2}{N(\rho)}\left(1\pm\sqrt{1-N(\rho)
\frac{1-\rho^2}{\rho^2}l_{12}}\right)
\end{equation}

As in the no-gravity case, we construct solution branches by fixing
$l_{12}$ and calculating $(k,\rho)$ pairs which may then be converted
via Eq.~\eqref{eq:quadratmittel} into $(k,\overline{\alpha})$ pairs.

Only the solutions for $l_{12}\in[0,1]$ cover the whole range
$\rho=0\dots 1$, while in the region $l_{12}>1$, i.e., below the
critical point, the system is linearly stable and hence we find no
solutions close to zero. In these cases it is necessary to first
calculate the minimum possible value of $\rho$ by requiring the
radicand in Eq.~\eqref{eq:k_extended} to be zero.

Clearly, the fact that finite-amplitude solutions exist at subcritical
values of $l_{12}$ is already indicative of the subcriticality of the
bifurcation at the threshold, a result first obtained by Nozi\`eres
\cite{nozieres:93}.  Let us discuss the vicinity of the critical point
in some more detail.  The neutral mode emerging at that point has of
course $\rho=0$.  So we should expand the energy or its derivative for
small $\rho$ to obtain the solution behavior at the bifurcation. Using
Eq.~\eqref{eq:tayln}, we find for the derivative of the total energy:
\begin{multline}
\label{eq:deretot} \frac{k}{2\pi}\frac{\partial E}{\partial\rho}
 = \left(-\frac1k
+\frac12 +\frac{l_{12}}{2k^2} \right) \rho
\\+\left(\frac{1}{16}-\frac{l_{12}}{2k^2}\right) \rho^3
 +\frac{3}{128} \rho^5  +
\frac{25}{2048} \rho^7 + \dots\>.
\end{multline}
At the bifurcation point, the linear term vanishes, because $k=1$ and
$l_{12}=1$. The third-order term is $-\nicefrac{7}{16}\,\rho^3$, i.e.,
it is negative, whereas higher-order terms are positive.  This is to be
contrasted with the Nozi\`eres calculation and its extension by
ourselves in \cite{kohlert:02a}, showing that all the calculable
coefficients of an amplitude expansion in terms of Fourier modes are
negative.  {From} this phenomenon we concluded that there is no
restabilization at finite amplitude.  At first sight, the situation
seems to be different here, due to the positivity of higher-order
coefficients. But in fact, it is not, because the maximum meaningful
amplitude is $\rho=1$, and it is easy to see that for this value the
third-order coefficient remains dominant. As $\rho\to1$, the energy
derivative tends to the negative value $4-2\pi$. In a sense, this
consideration shows a little more than the calculation in terms of
Fourier modes: restabilization of the structure would be possible at
$\rho>1$, but this is an unphysical situation.

\begin{figure}[htb]
\setlength{\unitlength}{\columnwidth}
\begin{picture}(1.000000,.510000)(-.150000,-.030000)
\allinethickness{.040000mm} \thinlines \thicklines
\drawline(0,0)(0,.480000)(.800000,.480000)(.800000,0)(0,0) \thinlines
\drawline(0.000000,.058537)(-.015000,.058537)
\put(-.025000,.058537){\makebox(0,0)[r]{-0.4}}
\drawline(0.000000,.175610)(-.015000,.175610)
\put(-.025000,.175610){\makebox(0,0)[r]{-0.2}}
\drawline(0.000000,.292683)(-.015000,.292683)
\put(-.025000,.292683){\makebox(0,0)[r]{0.0}}
\drawline(0.000000,.409756)(-.015000,.409756)
\put(-.025000,.409756){\makebox(0,0)[r]{0.2}}
\put(-.045000,.351220){\makebox(0,0)[r]{$\Im(\zeta)$}}
\drawline(0.000000,0.000000)(0.000000,-.015000)
\put(0.000000,-.025000){\makebox(0,0)[t]{0.00}}
\drawline(.200000,0.000000)(.200000,-.015000)
\put(.200000,-.025000){\makebox(0,0)[t]{2.42}}
\drawline(.400000,0.000000)(.400000,-.015000)
\put(.400000,-.025000){\makebox(0,0)[t]{4.83}}
\drawline(.600000,0.000000)(.600000,-.015000)
\put(.600000,-.025000){\makebox(0,0)[t]{7.25}}
\drawline(.800000,0.000000)(.800000,-.015000)
\put(.800000,-.025000){\makebox(0,0)[t]{9.67}}
\put(.700000,-.045000){\makebox(0,0)[t]{$\Re(\zeta)$}} \thinlines
\drawline(0.000000,.018012)(.004091,.019752)(.008243,.024943)
(.012517,.033505)(.016973,.045302)(.021666,.060147)(.026648,.077808)
(.031966,.098004)(.037664,.120419)(.043777,.144698)(.050335,.170458)
(.057360,.197294)(.064869,.224781)(.072869,.252487)(.081360,.279975)
(.090335,.306810)(.099777,.332571)(.109664,.356850)(.119966,.379264)
(.130648,.399461)(.141666,.417121)(.152973,.431967)(.164517,.443764)
(.176243,.452325)(.188091,.457517)(.200000,.459257)(.211909,.457517)
(.223757,.452325)(.235483,.443764)(.247027,.431967)(.258334,.417121)
(.269352,.399461)(.280034,.379264)(.290336,.356850)(.300223,.332571)
(.309665,.306810)(.318640,.279975)(.327131,.252487)(.335131,.224781)
(.342640,.197294)(.349665,.170458)(.356223,.144698)(.362336,.120419)
(.368034,.098004)(.373352,.077808)(.378334,.060147)(.383027,.045302)
(.387483,.033505)(.391757,.024943)(.395909,.019752)(.400000,.018012)
(.404091,.019752)(.408243,.024943)(.412517,.033505)(.416973,.045302)
(.421666,.060147)(.426648,.077808)(.431966,.098004)(.437664,.120419)
(.443777,.144698)(.450335,.170458)(.457360,.197294)(.464869,.224781)
(.472869,.252487)(.481360,.279975)(.490335,.306810)(.499777,.332571)
(.509664,.356850)(.519966,.379264)(.530648,.399461)(.541666,.417121)
(.552973,.431967)(.564517,.443764)(.576243,.452325)(.588091,.457517)
(.600000,.459257)(.611909,.457517)(.623757,.452325)(.635483,.443764)
(.647027,.431967)(.658334,.417121)(.669352,.399461)(.680034,.379264)
(.690336,.356850)(.700223,.332571)(.709665,.306810)(.718640,.279975)
(.727131,.252487)(.735131,.224781)(.742640,.197294)(.749665,.170458)
(.756223,.144698)(.762336,.120419)(.768034,.098004)(.773352,.077808)
(.778334,.060147)(.783027,.045302)(.787483,.033505)(.791757,.024943)
(.795909,.019752)(.800000,.018012)
\end{picture}
\caption{\label{fig:solexample} Two periods of a sample solution at $k=1.3$,
$\overline{\alpha} = 0.25$ (which corresponds to $\rho \approx 0.49$).}
\end{figure}
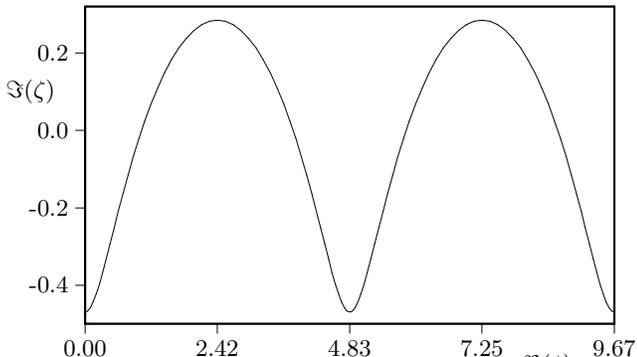

Figure \ref{fig:solexample} shows an example for the profile of a
steady state-solution taken from the critical branch ($l_{12}\!=\!1$).
The solution shows the typical behavior predicted by Nozi\`eres, i.e.,
flat cell tips and sharp grooves. It is however unstable, as shall be
seen from the discussion below.

\begin{figure}[htb]
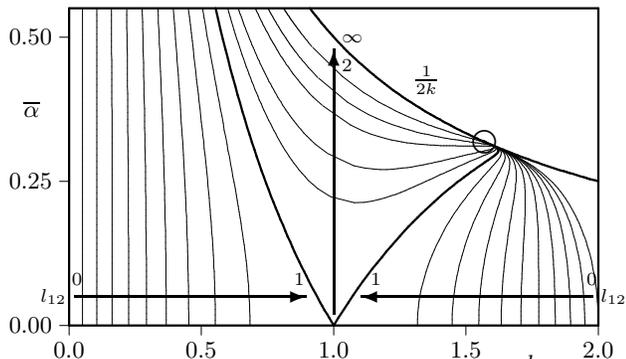

\setlength{\unitlength}{\columnwidth}

\caption{\label{fig:monozyk2}Solution branches of equation
  \eqref{eq:zweite} for different values of $l_{12}$. The rightmost
  curve, beginning at $(k,\bar{\alpha})=(2.0,0.0)$, is the no-gravity
  solution shown in Fig.~\ref{fig:monozyk}. Arrows and small numbers
  denote $l_{12}$ values corresponding to the curves. The upper thick
  line is the limiting curve for large $l_{12}$. The common
  termination point of all branches is marked by a circle. For more
  details of this region see Fig.~\ref{fig:monozyk3}.}
\end{figure}

A number of solution branches for different values of $l_{12}$ is
shown in Fig.~\ref{fig:monozyk2}. In this plot, we first of all see
how the range of unstable wave numbers, starting with the interval
$(0,2)$ becomes narrower with increasing $l_{12}$, i.e., higher
gravity or lower prestress. At $l_{12}=1$, only a single $k$ value has
a non-negative growth rate; this is the critical case.

At even higher values of $l_{12}$, we are in the subcritical range
where it takes some energy to modulate the surface in order to
overcome the energy barrier to let the instability emerge.  In
Fig.~\ref{fig:monozyk2}, we have shown solutions up to $l_{12}=2$. The
solution branches converge towards $\overline\alpha=\nicefrac{1}{2k}$
as $l_{12}$ is increased, which is the transformation of the cusp
limit $\rho=1$ (see equation \eqref{eq:quadratmittel}). The common
endpoint of \textsl{all} curves
$(\nicefrac{\pi}{2},\nicefrac{1}{\pi})$ is marked with a circle.

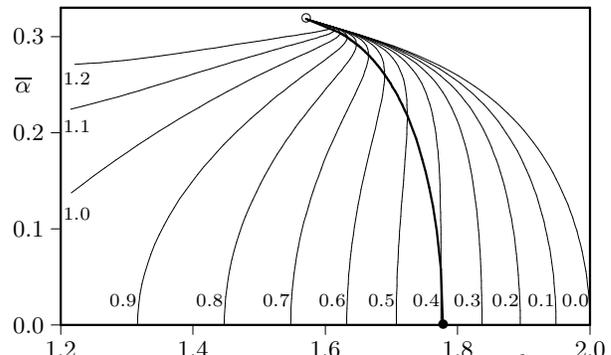
\begin{figure}[htb]
\setlength{\unitlength}{\columnwidth}
\begin{picture}(1.000000,.510000)(-.150000,-.030000)
\allinethickness{.040000mm} \thinlines \thicklines
\drawline(0,0)(0,.480000)(.800000,.480000)(.800000,0)(0,0) \thinlines
\drawline(0.000000,0.000000)(-.015000,0.000000)
\put(-.025000,0.000000){\makebox(0,0)[r]{0.0}}
\drawline(0.000000,.145455)(-.015000,.145455)
\put(-.025000,.145455){\makebox(0,0)[r]{0.1}}
\drawline(0.000000,.290909)(-.015000,.290909)
\put(-.025000,.290909){\makebox(0,0)[r]{0.2}}
\drawline(0.000000,.436364)(-.015000,.436364)
\put(-.025000,.436364){\makebox(0,0)[r]{0.3}}
\put(-.045000,.363636){\makebox(0,0)[r]{$\overline{\alpha}$}}
\drawline(0.000000,0.000000)(0.000000,-.015000)
\put(0.000000,-.025000){\makebox(0,0)[t]{1.2}}
\drawline(.200000,0.000000)(.200000,-.015000)
\put(.200000,-.025000){\makebox(0,0)[t]{1.4}}
\drawline(.400000,0.000000)(.400000,-.015000)
\put(.400000,-.025000){\makebox(0,0)[t]{1.6}}
\drawline(.600000,0.000000)(.600000,-.015000)
\put(.600000,-.025000){\makebox(0,0)[t]{1.8}}
\drawline(.800000,0.000000)(.800000,-.015000)
\put(.800000,-.025000){\makebox(0,0)[t]{2.0}}
\put(.700000,-.045000){\makebox(0,0)[t]{$k$}} \thinlines
\drawline(.800000,0.000000)(.799903,.010285)(.799600,.020566)
(.799100,.030842)(.798397,.041108)(.797494,.051362)(.796387,.061600)
(.795076,.071820)(.793559,.082017)(.791833,.092190)(.789898,.102335)
(.787750,.112448)(.785387,.122527)(.782805,.132568)(.780001,.142568)
(.776971,.152523)(.773711,.162430)(.770216,.172286)(.766481,.182086)
(.762500,.191829)(.758267,.201509)(.753775,.211124)(.749017,.220669)
(.743984,.230141)(.738667,.239537)(.733055,.248853)(.727139,.258084)
(.720904,.267227)(.714339,.276278)(.707427,.285234)(.700151,.294091)
(.692492,.302846)(.684429,.311495)(.675938,.320034)(.666989,.328463)
(.657552,.336777)(.647590,.344977)(.637059,.353060)(.625910,.361029)
(.614080,.368886)(.601498,.376637)(.588074,.384290)(.573695,.391861)
(.558217,.399372)(.541449,.406859)(.523128,.414377)(.502872,.422018)
(.480080,.429937)(.453689,.438434)(.421338,.448211)(.370796,.462996)

\drawline(.748683,0.000000)(.748607,.010555)(.748367,.021107)
(.747972,.031651)(.747417,.042184)(.746702,.052702)(.745827,.063200)
(.744788,.073677)(.743586,.084126)(.742217,.094545)(.740681,.104930)
(.738973,.115277)(.737092,.125582)(.735034,.135841)(.732796,.146050)
(.730373,.156205)(.727761,.166302)(.724957,.176336)(.721953,.186305)
(.718745,.196203)(.715326,.206027)(.711689,.215772)(.707827,.225433)
(.703730,.235008)(.699391,.244490)(.694798,.253877)(.689941,.263163)
(.684806,.272345)(.679380,.281417)(.673648,.290377)(.667593,.299218)
(.661195,.307939)(.654432,.316533)(.647280,.324999)(.639711,.333333)
(.631692,.341532)(.623188,.349594)(.614154,.357518)(.604539,.365305)
(.594284,.372956)(.583314,.380477)(.571541,.387877)(.558850,.395168)
(.545099,.402374)(.530095,.409529)(.513575,.416687)(.495157,.423939)
(.474239,.431437)(.449759,.439479)(.419356,.448760)(.370796,.462996)

\drawline(.694427,0.000000)(.694375,.010858)(.694205,.021711)
(.693926,.032554)(.693535,.043384)(.693031,.054196)(.692412,.064984)
(.691677,.075745)(.690824,.086474)(.689852,.097165)(.688757,.107815)
(.687538,.118419)(.686191,.128971)(.684712,.139468)(.683099,.149904)
(.681346,.160275)(.679449,.170576)(.677404,.180803)(.675204,.190949)
(.672844,.201012)(.670317,.210985)(.667616,.220863)(.664732,.230643)
(.661657,.240319)(.658382,.249886)(.654895,.259339)(.651185,.268673)
(.647239,.277883)(.643043,.286966)(.638582,.295915)(.633836,.304726)
(.628788,.313395)(.623414,.321918)(.617689,.330290)(.611585,.338508)
(.605070,.346569)(.598105,.354471)(.590647,.362211)(.582644,.369792)
(.574036,.377213)(.564748,.384480)(.554690,.391601)(.543749,.398590)
(.531779,.405469)(.518589,.412271)(.503914,.419050)(.487370,.425895)
(.468358,.432958)(.445811,.440533)(.417369,.449311)(.370796,.462996)

\drawline(.636660,0.000000)(.636634,.011199)(.636544,.022392)
(.636398,.033574)(.636192,.044739)(.635925,.055882)(.635597,.066996)
(.635205,.078076)(.634747,.089117)(.634222,.100112)(.633626,.111057)
(.632956,.121945)(.632209,.132771)(.631381,.143529)(.630468,.154214)
(.629465,.164820)(.628368,.175342)(.627171,.185773)(.625866,.196109)
(.624449,.206344)(.622912,.216471)(.621246,.226487)(.619443,.236384)
(.617493,.246158)(.615387,.255804)(.613113,.265315)(.610658,.274686)
(.608009,.283913)(.605151,.292990)(.602067,.301911)(.598738,.310672)
(.595145,.319269)(.591263,.327696)(.587067,.335950)(.582528,.344026)
(.577612,.351922)(.572280,.359636)(.566488,.367165)(.560184,.374510)
(.553304,.381673)(.545774,.388659)(.537503,.395475)(.528377,.402136)
(.518248,.408662)(.506924,.415088)(.494140,.421467)(.479511,.427888)
(.462434,.434501)(.441844,.441597)(.415376,.449866)(.370796,.462996)

\drawline(.574597,0.000000)(.574601,.011590)(.574603,.023174)
(.574611,.034743)(.574619,.046291)(.574628,.057812)(.574636,.069297)
(.574640,.080741)(.574638,.092135)(.574627,.103474)(.574605,.114750)
(.574566,.125957)(.574508,.137088)(.574426,.148136)(.574314,.159095)
(.574167,.169958)(.573979,.180718)(.573744,.191369)(.573454,.201905)
(.573102,.212319)(.572679,.222606)(.572176,.232758)(.571584,.242770)
(.570892,.252636)(.570088,.262350)(.569159,.271906)(.568093,.281299)
(.566873,.290523)(.565484,.299572)(.563908,.308442)(.562125,.317127)
(.560112,.325623)(.557846,.333925)(.555299,.342030)(.552441,.349933)
(.549236,.357631)(.545645,.365123)(.541621,.372408)(.537111,.379485)
(.532051,.386357)(.526364,.393029)(.519959,.399509)(.512718,.405812)
(.504495,.411959)(.495095,.417985)(.484251,.423942)(.471576,.429919)
(.456468,.436066)(.437857,.442672)(.413379,.450422)(.370796,.462996)

\drawline(.507107,0.000000)(.507148,.012048)(.507260,.024088)
(.507451,.036110)(.507716,.048105)(.508053,.060065)(.508461,.071981)
(.508937,.083845)(.509477,.095647)(.510077,.107380)(.510734,.119035)
(.511442,.130603)(.512195,.142077)(.512989,.153449)(.513816,.164711)
(.514669,.175855)(.515541,.186874)(.516424,.197760)(.517309,.208506)
(.518185,.219105)(.519044,.229551)(.519873,.239836)(.520662,.249955)
(.521398,.259900)(.522067,.269666)(.522654,.279247)(.523145,.288637)
(.523522,.297830)(.523767,.306822)(.523861,.315608)(.523780,.324182)
(.523502,.332540)(.523000,.340679)(.522246,.348594)(.521206,.356283)
(.519843,.363743)(.518117,.370973)(.515979,.377972)(.513374,.384742)
(.510236,.391285)(.506487,.397607)(.502034,.403717)(.496757,.409629)
(.490509,.415367)(.483093,.420965)(.474241,.426477)(.463564,.431989)
(.450457,.437654)(.433851,.443758)(.411377,.450982)(.370796,.462996)

\drawline(.432456,0.000000)(.432543,.012599)(.432793,.025186)
(.433213,.037751)(.433798,.050281)(.434545,.062766)(.435452,.075195)
(.436514,.087555)(.437725,.099838)(.439081,.112031)(.440575,.124125)
(.442200,.136110)(.443947,.147976)(.445810,.159713)(.447777,.171312)
(.449841,.182765)(.451989,.194063)(.454212,.205197)(.456497,.216161)
(.458832,.226945)(.461204,.237544)(.463598,.247949)(.466000,.258156)
(.468395,.268157)(.470766,.277946)(.473095,.287518)(.475365,.296868)
(.477556,.305991)(.479646,.314882)(.481613,.323537)(.483434,.331951)
(.485080,.340122)(.486525,.348047)(.487736,.355722)(.488679,.363146)
(.489315,.370317)(.489600,.377234)(.489484,.383900)(.488910,.390315)
(.487811,.396484)(.486107,.402413)(.483701,.408112)(.480475,.413598)
(.476278,.418894)(.470912,.424034)(.464106,.429074)(.455472,.434101)
(.444402,.439265)(.429824,.444854)(.409370,.451545)(.370796,.462996)

\drawline(.347723,0.000000)(.347872,.013288)(.348307,.026561)
(.349035,.039803)(.350048,.052998)(.351344,.066133)(.352916,.079191)
(.354756,.092159)(.356856,.105023)(.359207,.117770)(.361798,.130386)
(.364618,.142859)(.367655,.155177)(.370894,.167329)(.374322,.179305)
(.377925,.191095)(.381686,.202689)(.385589,.214078)(.389618,.225255)
(.393754,.236212)(.397981,.246942)(.402279,.257438)(.406629,.267696)
(.411011,.277708)(.415404,.287472)(.419788,.296981)(.424138,.306232)
(.428434,.315221)(.432649,.323946)(.436758,.332403)(.440734,.340590)
(.444548,.348505)(.448168,.356147)(.451560,.363514)(.454688,.370605)
(.457509,.377423)(.459980,.383966)(.462046,.390238)(.463650,.396241)
(.464723,.401982)(.465183,.407469)(.464933,.412713)(.463852,.417730)
(.461788,.422546)(.458542,.427197)(.453842,.431737)(.447298,.436255)
(.438301,.440901)(.425778,.445961)(.407358,.452110)(.370796,.462996)

\drawline(.247214,0.000000)(.247458,.014210)(.248174,.028397)
(.249370,.042540)(.251034,.056615)(.253156,.070601)(.255725,.084478)
(.258724,.098226)(.262139,.111827)(.265949,.125262)(.270136,.138515)
(.274677,.151572)(.279549,.164418)(.284730,.177040)(.290194,.189428)
(.295917,.201571)(.301873,.213460)(.308037,.225087)(.314383,.236446)
(.320885,.247529)(.327518,.258333)(.334255,.268852)(.341070,.279084)
(.347938,.289024)(.354831,.298671)(.361724,.308022)(.368590,.317077)
(.375400,.325833)(.382127,.334291)(.388742,.342449)(.395215,.350309)
(.401513,.357870)(.407603,.365134)(.413449,.372100)(.419013,.378772)
(.424251,.385151)(.429116,.391240)(.433556,.397044)(.437510,.402567)
(.440909,.407816)(.443671,.412802)(.445697,.417537)(.446866,.422039)
(.447025,.426334)(.445974,.430458)(.443443,.434469)(.439039,.438453)
(.432152,.442562)(.421710,.447080)(.405340,.452678)(.370796,.462996)

\drawline(.116228,0.000000)(.116664,.015622)(.117950,.031203)
(.120086,.046706)(.123042,.062092)(.126789,.077325)(.131289,.092374)
(.136498,.107210)(.142371,.121804)(.148856,.136136)(.155903,.150185)
(.163460,.163935)(.171478,.177374)(.179904,.190489)(.188692,.203274)
(.197795,.215721)(.207168,.227826)(.216769,.239587)(.226558,.251002)
(.236497,.262071)(.246550,.272793)(.256681,.283170)(.266858,.293203)
(.277048,.302896)(.287221,.312249)(.297347,.321265)(.307396,.329948)
(.317339,.338301)(.327145,.346326)(.336786,.354027)(.346230,.361407)
(.355445,.368469)(.364399,.375217)(.373055,.381655)(.381376,.387787)
(.389319,.393616)(.396839,.399148)(.403882,.404389)(.410391,.409346)
(.416296,.414026)(.421517,.418442)(.425955,.422607)(.429491,.426539)
(.431972,.430266)(.433200,.433825)(.432904,.437273)(.430692,.440698)
(.425956,.444249)(.417622,.448210)(.403318,.453249)(.370796,.462996)

\drawline(.016135,.200031)(.033260,.213140)(.050225,.225787)
(.067021,.237986)(.083642,.249750)(.100082,.261091)(.116331,.272021)
(.132383,.282549)(.148228,.292687)(.163857,.302442)(.179259,.311825)
(.194425,.320843)(.209342,.329503)(.223999,.337813)(.238380,.345780)
(.252472,.353409)(.266259,.360707)(.279722,.367680)(.292841,.374332)
(.305596,.380669)(.317962,.386696)(.329911,.392418)(.341414,.397840)
(.352434,.402968)(.362933,.407807)(.372864,.412364)(.382174,.416646)
(.390798,.420663)(.398660,.424425)(.405666,.427947)(.411699,.431248)
(.416612,.434354)(.420208,.437304)(.422218,.440153)(.422255,.442990)
(.419710,.445962)(.413513,.449351)(.401290,.453823)(.370796,.462996)

\drawline(.014935,.326513)(.034453,.331195)(.052854,.335928)
(.070342,.340665)(.087063,.345371)(.103124,.350025)(.118606,.354609)
(.133571,.359111)(.148069,.363522)(.162139,.367834)(.175810,.372042)
(.189108,.376143)(.202053,.380133)(.214659,.384010)(.226939,.387772)
(.238902,.391417)(.250555,.394946)(.261901,.398357)(.272943,.401650)
(.283682,.404825)(.294115,.407882)(.304239,.410822)(.314050,.413645)
(.323540,.416352)(.332700,.418945)(.341520,.421423)(.349985,.423790)
(.358081,.426048)(.365789,.428198)(.373086,.430245)(.379947,.432191)
(.386340,.434043)(.392230,.435805)(.397571,.437487)(.402310,.439097)
(.406381,.440649)(.409702,.442160)(.412165,.443654)(.413631,.445162)
(.413908,.446732)(.412716,.448434)(.409621,.450386)(.403860,.452800)
(.393756,.456160)(.370805,.462994)
\drawline(.021415,.394323)(.041546,.395415)(.060376,.396791)
(.078139,.398367)(.095004,.400083)(.111098,.401896)(.126517,.403775)
(.141335,.405694)(.155611,.407634)(.169394,.409580)(.182721,.411521)
(.195624,.413445)(.208129,.415345)(.220256,.417215)(.232023,.419048)
(.243443,.420840)(.254527,.422587)(.265284,.424287)(.275718,.425936)
(.285835,.427533)(.295636,.429076)(.305121,.430564)(.314289,.431995)
(.323136,.433370)(.331657,.434689)(.339844,.435951)(.347687,.437159)
(.355176,.438312)(.362295,.439413)(.369027,.440464)(.375351,.441468)
(.381241,.442431)(.386668,.443357)(.391593,.444253)(.395972,.445129)
(.399749,.445996)(.402853,.446869)(.405195,.447771)(.406654,.448730)
(.407066,.449788)(.406193,.451007)(.403664,.452490)(.398830,.454420)
(.390266,.457211)(.370805,.462994) \thicklines
\drawline(.577778,0.000000)(.574640,.083724)(.568504,.142351)
(.562644,.179923)(.557043,.208465)(.551684,.231574)(.546552,.250943)
(.541634,.267541)(.536917,.281985)(.532389,.294702)(.528041,.306000)
(.523861,.316112)(.519841,.325218)(.515972,.333461)(.512246,.340955)
(.508656,.347797)(.505195,.354064)(.501856,.359823)(.498634,.365130)
(.495522,.370033)(.492516,.374574)(.489611,.378790)(.486802,.382710)
(.484084,.386363)(.481454,.389773)(.478907,.392962)(.476441,.395949)
(.474051,.398750)(.471734,.401381)(.469487,.403856)(.467308,.406187)
(.465194,.408384)(.463142,.410459)(.461149,.412419)(.459213,.414274)
(.457333,.416031)(.455505,.417696)(.453729,.419276)(.452001,.420777)
(.450321,.422204)(.448686,.423561)(.447095,.424854)(.445547,.426085)
(.444039,.427260)(.442570,.428381)(.441140,.429452)(.439746,.430476)
(.438388,.431455)(.437064,.432392)(.435773,.433290)(.434514,.434150)
(.433287,.434974)(.432089,.435766)(.430921,.436525)(.429780,.437255)
(.428667,.437956)(.427580,.438630)(.426519,.439279)(.425483,.439904)
(.424470,.440505)(.423482,.441085)(.422515,.441643)(.370796,.462996)

\put(.370796,.462996){\makebox(0,0){$\circ$}}
\put(.577778,0.000000){\makebox(0,0){$\bullet$}}
\put(.798000,.029091){\makebox(0,0)[rb]{{\scriptsize 0.0}}} \thinlines
\put(.746683,.029091){\makebox(0,0)[rb]{{\scriptsize 0.1}}} \thinlines
\put(.692427,.029091){\makebox(0,0)[rb]{{\scriptsize 0.2}}} \thinlines
\put(.634660,.029091){\makebox(0,0)[rb]{{\scriptsize 0.3}}} \thinlines
\put(.572597,.029091){\makebox(0,0)[rb]{{\scriptsize 0.4}}} \thinlines
\put(.505107,.029091){\makebox(0,0)[rb]{{\scriptsize 0.5}}} \thinlines
\put(.430456,.029091){\makebox(0,0)[rb]{{\scriptsize 0.6}}} \thinlines
\put(.345723,.029091){\makebox(0,0)[rb]{{\scriptsize 0.7}}} \thinlines
\put(.245214,.029091){\makebox(0,0)[rb]{{\scriptsize 0.8}}} \thinlines
\put(.114228,.029091){\makebox(0,0)[rb]{{\scriptsize 0.9}}} \thinlines
\put(.004000,.160000){\makebox(0,0)[lb]{{\scriptsize 1.0}}} \thinlines
\put(.004000,.290909){\makebox(0,0)[lb]{{\scriptsize 1.1}}} \thinlines
\put(.004000,.363636){\makebox(0,0)[lb]{{\scriptsize 1.2}}} \thinlines
\end{picture}
\caption{\label{fig:monozyk3}
  Solution branches of equation \eqref{eq:zweite} in a certain
  $k$-range. The region bounded by the no-gravity solution and the
  curve (thick line) from the bullet symbol up to the cusp point
  contains the stable solutions.}
\end{figure}

Again, as in the no-gravity case we have to check the stability of the
solutions by inserting the corresponding solution
\eqref{eq:k_extended} into the second derivative of the energy. First
we note that all branches left of $k=\nicefrac{\pi}{2}$ in
Fig.~\ref{fig:monozyk2} start off unstably. But there is a range of
stability which we discuss with the help of Fig.~\ref{fig:monozyk3}
rather than \ref{fig:monozyk2}.

In this figure, we have included the range of stable amplitude
solutions. The largest $l_{12}$ value where \textsl{small} solutions
still emerge stably ($\bullet$) is $l_{12}=\nicefrac{32}{81}$,
corresponding to $k=\nicefrac{16}{9}$. This should be compared with
the result of \cite{kohlert:02a}, where the tricritical point is shown
to be located at $k=\nicefrac{13}{3}-\nicefrac{\sqrt{57}}{3}\approx
1.82$, corresponding to
$l_{12}=\nicefrac{20\sqrt{57}}{9}-\nicefrac{148}{9}\approx 0.33$.
Because the result from the amplitude equations is exact at
infinitesimal amplitudes, the variational ansatz overestimates the
range of supercritical bifurcation at lowest order. Of course, such a
comparison is a bit unfair towards the variational approach, as we
have only a single para\-meter available there, whereas the
lowest-order nonlinear mode expansion contains already two amplitudes.
As soon as we take the multi-cycloid ansatz, discussed below, to
second order, we obtain the exact position of the tricritical point
within this ansatz as well. This is simply due to the fact that the
$n$-th summand in the multi-cycloid expression
(Eq.~\ref{eq:multigenerator}) does not contain Fourier modes lower
than $n$, hence all contributions of order 2 must be present for $N=2$
(but some are already present for $N=1$).

An interesting fact is the bending of all solution branches into a
stable region before running into the cusp, a feature not obtained
within the amplitude equation approach.  Actually, the stability
changes at the points where the slope turns negative (passing from
$+\infty$ to $-\infty$).  This can be interpreted as a hint for the
existence of stable solution branches at high gravity. These stable
solutions do not, at least in the mono-cycloid approximation, emerge at
the upper marginal wave number $1+\sqrt{1-l_{12}}$ predicted by the
linear stability analysis but instead in the (upper) vicinity of
$k=\nicefrac{\pi}{2}$ at nonzero amplitude. Whether this is really
true, will be checked in the following chapters by the implementation
of the multi-cycloid ansatz.

\section{\label{sec:multicyc}The multi-cycloid approximation}

In seeking a generalization of the cycloid ansatz, we were guided by
two principles. First, the generalization should reduce to the
cycloid, when all but one of the parameters became small. Second, it
should correspond to a boundary curve that allows us the analytic
solution of the elastic problem by conformal mapping.

Given these conditions, the multi-cycloid model is a natural
generalization of the cycloid one (and of a two-cycloid ansatz already
considered in \cite{kassner:01}):
%from ansatz \eqref{eq:monogenerator} :
\begin{equation}
\label{eq:multigenerator} \zeta(\xi)=\xi-i\sum_{n=1}^{N}\frac{\rho_n}{nk}
e^{-ink\xi}\>.
\end{equation}
Herein, $N$ is the number of ''cycloid modes'' taken into account. The
denominator $n$ in this equation which could also have been included
into the definition of the $\rho_n$ has been explicitly written in
order to be able to express the generalized cusp condition in a compact
way. Real and imaginary parts read:
\begin{subequations}
\label{eq:multigenerator2}
\begin{align}
\label{eq:multigenerator_re} x(\xi)&= \xi-\sum_{n=1}^{N}\frac{\rho_n}{nk}
\sin{(nk\xi)}
\>,\\ \label{eq:multigenerator_im} y(\xi)&= -\sum_{n=1}^{N}\frac{\rho_n}{nk}
\cos{(nk\xi)} \>.
\end{align}
\end{subequations}
Again we shift the interface by its mean value to set the average
interface position equal to zero:
\begin{equation}
m=\frac{k}{2\pi}\int\dot{x}(\xi)y(\xi){\rm
d}\xi=\frac{1}{2k}\sum_{n=1}^{N}\frac{\rho_n^2}{n}\>,
\end{equation}
and we have to correct Eq.~\eqref{eq:multigenerator_im} as follows:
\begin{equation}
\label{eq:multigenerator_im2} y(\xi)=-\sum_{n=1}^{N}\frac{\rho_n}{nk}
\left(\cos{(nk\xi)}+\frac{\rho_n}{2}\right) \>.
\end{equation}
Moreover, we will need the mean square amplitude again. The result is
\begin{equation}
\label{eq:multiquadratmittel} \overline{\alpha}=\left|\frac{1}{k}
\sqrt{\sum_{n=1}^{N}\left[ \!\frac{\rho_n^{\,2}}{2n^2}\!\!
\left[\!1\!-\!\frac{n}{2}\sum_{j=1}^{N}\frac{\rho_j^{\,2}}{j}\!\right]\!\!
-\!\!\sum_{j=1}^{N-n}\frac{\rho_n\rho_j\rho_{n+j}}{j\left(n\!+\!j\right)}
\right]}\right|.
\end{equation}
We assume the sequence of the $\rho_n$ to decrease sufficiently fast
so the sum of the absolute values of the $\rho_n$ does not exceed one,
a condition that is sufficient to avoid self-crossings of the curve
given by \eqref{eq:multigenerator}. Then cusps, if they exist, can
appear only at
\begin{equation}
\label{eq:cusp_at}\xi_{\rm cusp}\,k=2\pi{n},~n\!\in\!\mathbbm{N}\>.
\end{equation}
(They are characterized by $\zeta'(\xi)=0$.) The radius of curvature is
given as \cite{bartsch:1}
\begin{equation}
r=\left|\frac{\left(\dot{x}(\xi)^2+\dot{y}(\xi)^2\right)^\frac{3}{2}}{
\left|\begin{array}{cc}\dot{x}(\xi)&\dot{y}(\xi)\\
\ddot{x}(\xi)&\ddot{y}(\xi)\\
\end{array}\right|}\right|\>,
\end{equation}
and it takes its minimum value at $\xi\!=\!\xi_{\rm cusp}$. Therefore,
\begin{equation}
r_{\xi=0}=\left|\frac{\left(-1+\sum\limits_{n=1}^N\rho_n\right)^2}
{k\sum\limits_{n=1}^N
n\rho_n }\right|\>,
\end{equation}
and the cusp condition reads
\begin{equation}
\label{eq:generalized_cusp} \sum_{n=1}^{N}\rho_n=1\>.
\end{equation}

Now the derivatives of the relevant energies have to be calculated.
Note that the integration for the energies themselves may not be
carried out analytically for general multi-cycloids. But the question
of stability of the solution can be answered, because first and second
derivative can be given explicitely.

The crucial point is the calculation of the elastic energy density
$w(s)$ at the surface, which is done in appendix
\ref{app:generalmapping}. Supposing $w(s)$ is given (see
Eq.~\eqref{eq:diewirklicheenergiedichte}), we have to modify
Eqs.~\eqref{eq:dx} and \eqref{eq:nds} as follows (we write $C_n$ and
$S_n$ instead of $\cos(nk\xi)$ and $\sin(nk\xi)$):

\begin{align}
\mathbf{\delta x}=&\sum_{n=1}^{N}\left(\frac{\partial x(\xi)}{\partial
\rho_n}\mathbf{e}_x+\frac{\partial y(\xi)}{\partial
\rho_n}\mathbf{e}_y\right)\delta\rho_n\nonumber\\=&-\frac{1}{nk}\sum_{n=1}^{N}
\left[S_n\mathbf{e}_x+\left(C_n+\rho_n\right)\mathbf{e}_y\right]\delta\rho_n\>,
\label{eq:multi_dx}\\
\mathbf{n}ds=& \left(-\dot{y}\left(\xi\right)\mathbf{e}_x+\dot{x}
\left(\xi\right)\mathbf{e}_y\right)d\xi \label{eq:multi_nds}
\end{align}

Consequently, we get according to Eq.~\eqref{eq:du}
\begin{equation}\label{eq:dEen}\frac{\partial E_{\rm
e}}{\partial\rho_n}=\frac{1}{nk}\int w(s) \left[ S_n
\dot{y}(\xi)-\left(C_n+\rho_n\right)\dot{x}(\xi) \right] d\xi\>. \end{equation}

The other terms are simpler again. Gravitational energy is $l_{12}/2$
times the square of the mean square amplitude
\eqref{eq:multiquadratmittel}
\begin{equation}
E_{\rm g}=\frac{l_{12}}{2}\,{\overline\alpha}^2\>,
\end{equation}
if we divide the integral by the
wavelength, as we will do in all energy expressions from now on, i.e.,
rather than integrals over a periodicity unit we consider averages.
We obtain
\begin{equation} \label{eq:dEgn}
\frac{\partial E_{\rm g}}{\partial\rho_n}\!=\!\frac{l_{12}}{2k^2}
\left[\frac{\rho_n}{n}\!\left(\!\!\frac{1}{n}
-\!\!\sum_{j=1}^{N}\frac{\rho_j^2}{j}\right) -\!\!\!\!\sum_{j=1,j\neq{n}}^{N}
\frac{\rho_{j}\rho_{\left|n-j\right|}}{n\left|n-j\right|}\right]\!,
\end{equation}
and the surface tension is again represented by the difference of the
arc lengths (compare Eq.~\eqref{eq:bogenlaenge}), hence its derivative
reads after some simplification
\begin{multline}
\label{eq:dEsn}\frac{\partial E_{\rm s}}{\partial\rho_n}= \int\mathrm{d}\xi\left[\frac{
\rho_{n}-C_n +\sum\limits_{l=1}^{n-1}C_l\rho_{n-l}+
\sum\limits_{l=1}^{N-n}C_l\rho_{n+l}}{
\sqrt{1+f(\overline{\rho},\xi)}}\right]\>,\\
f(\overline{\rho},\xi)=\sum\limits _{l=1}^{N}\rho_l\left(\rho_l-2C_l\right)
+2\sum\limits_{l=1}^{N-1}C_l\sum\limits _{j=1}^{N-l}\rho_{{j}}\rho_{{j+l}}
\end{multline}
This integral can only be solved analytically for the case $N\!=\!1$;
the analytical result has been used in the mono-cycloid model.

Equipped with the terms \eqref{eq:dEen}, \eqref{eq:dEgn} and
\eqref{eq:dEsn} we can now numerically solve the system
\begin{equation}
\label{eq:dritte}\frac{\partial}{\partial\rho_n}\left(E_{\rm e}
 + E_{\rm s} + E_{\rm g}\right)=0\>,\quad n=1\dots N\>,
\end{equation}
for the set $\{ \rho_1 \dots \rho_N \}$, given a certain prescribed
$k$ value. Some technical details of the solution method are described
in appendix \ref{app:howtosolve}.

\begin{figure}[htb]
\resizebox{0.95\columnwidth}{!}{\includegraphics{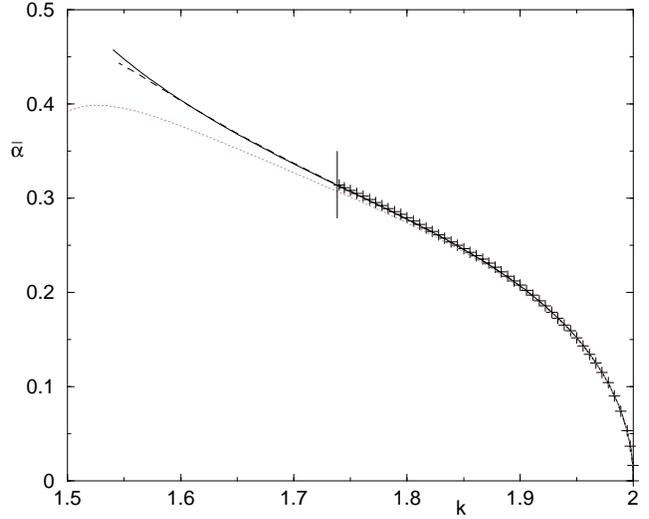}}
\caption{\label{fig:multivergleich}
  Comparison of the $N=2, 3$ and $10$ multi-cycloid models in the
  no-gravity case. The dotted line is the $N=2$ approximation, which
  is insufficient for our purpose: it does not reach the cusp at all
  but turns into a bag-like morphology instead (i.e., it develops
  overhangs). The dashed line is $N=3$, being already in good
  agreement with \cite{spencer:94}, and the solid line is the $N=10$
  example. All solutions with $N\geq4$ look the same.  Curves are
  terminating slightly before the cusp emerges, which will be further
  clarified in fig.\ref{fig:kruemmungen}. The crosses represent the
  numerical solution from \cite{spencer:94}.}
\end{figure}

We have carried out the calculations for a range of $N$ up to 50.
Figure \ref{fig:multivergleich} shows how fast the effective amplitude
branches converge to the numerical result from \cite{spencer:94}.

\begin{figure}[htb]
\resizebox{0.95\columnwidth}{!}{\includegraphics{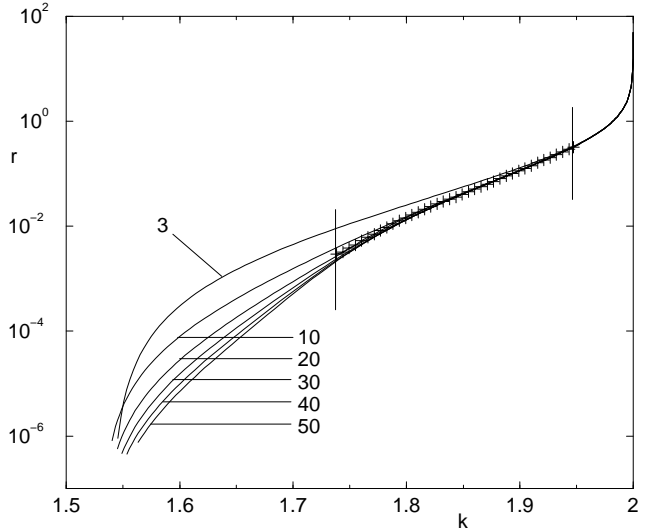}}
\caption{\label{fig:kruemmungen}
  Radii of curvature for $N$ up to 50, explanations see text.  Again,
  the crosses show the digitized and scaled radii of curvature of the
  Spencer solution.}
\end{figure}

A means to assess the closeness of a solution branch to its cusp
termination is to determine the radius of curvature in the minimum of
the grooves which will approach zero near the cusp. Here we note that
the $N=3$ approximation is insufficient for a description of the tip
radius (Fig.~\ref{fig:kruemmungen}), in contrast with the effective
amplitude (Fig.~\ref{fig:multivergleich}), which is already well
approximated by three modes for most of the branch.  (We have omitted
$N=2$ here because it is \textsl{much} worse.)

As can be seen, the solutions agree with the Spencer solution in the
range they cover.  Yet, the extension to the range of lower
wavenumbers is rather sensitive to the number of included modes.  A
prolongation of the curves in Fig. \ref{fig:kruemmungen} suggests a
termination point slightly left of $k=1.55$, which is less than
$\nicefrac{\pi}{2}$ as proposed by the monocycloid model. As with
increasing $N$ the termination point moves to slightly higher values
of $k$, it is tempting to speculate that the exact termination point
is at $k=\pi/2$ indeed. In any case, our method allows to reach radii
of curvature that are three orders of magnitude smaller than the
minimum value found by Spencer and Meiron, and it does so apparently
with less numerical effort.

\begin{figure}[htb]
\resizebox{0.95\columnwidth}{!}{\includegraphics{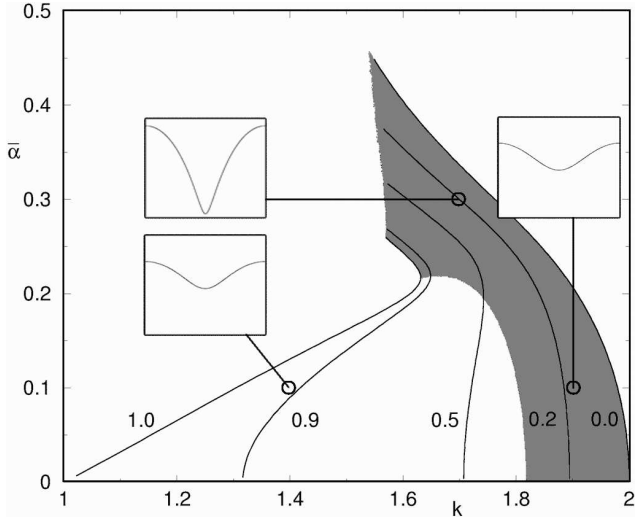}}
\caption{\label{fig:multizyk}
  Solution branches based on a $N\!=\!30$ multi-cycloid approximation.
  Numbers denote the value of $l_{12}$. The shaded area indicates the
  stable part of the solution manifold. Small sub-figures represent
  example morphologies.}
\end{figure}

The incorporation of gravity has been carried out up to $l_{12}=1$.
Figure \ref{fig:multizyk} gives the solution branches, unstable curves
left of $k=1$ are not shown. We note that the common cusp point, an
important feature of Fig.~\ref{fig:monozyk3}, disappears. As in
Fig.~\ref{fig:multivergleich}, the curves terminate before the cusp is
actually reached.

Stability of the solutions is checked via computation of the
determinant of the matrix $\partial^2 E/\partial \rho_n
\partial \rho_m$ and its principal minors. If all of these are
positive, the matrix is positive definite, the energy has a minimum
and the solution is stable. As it turns out, the determinant itself
gets positive only after its minors, so it would in our case be
sufficient to check the sign of the determinant alone.  The range of
stable solutions is displayed as the grayed area in
Fig.~\ref{fig:multizyk}.  We conclude that the stability behavior is
qualitatively correctly described by the mono-cycloid model already.
However, stability sets in at smaller amplitudes for $k$ values below
$1.7$. Hence, the multi-cycloid modes act stabilizing at larger
amplitudes and destabilizing near the tricritical point, which is
shifted to larger $k$ by inclusion of the second mode.

The subfigures of \ref{fig:multizyk} show that up to the different
distances to the cusp, which make the curves more or less ''sharp'',
there is no indication from the shape itself whether it is stable or
not.

\section{\label{sec:summary}Summary}
To conclude, we have presented a variational approach to the
calculation of steady states for the Grinfeld instability. Taking into
account a single mode we already obtain a very nice qualitative
description of the system behavior including the approach to a cusped
state. The wavenumber for the cusp appearance is already more accurate
with a single mode than in the article of Spencer and Meiron, while
the amplitude is pretty far off the true result (by about the same
amount as the amplitude obtained by Spencer and Meiron).

Nevertheless, this single-mode approximation has the virtue of great
transparency. That a cusp singularity appears is rendered
understandable: the system simply draws near a state where further
minimization of the elastic energy would require the interface to
self-intersect.

A few words may be in order concerning the limits of validity of our
approach. The nature of our calculation is variational, which means
that it will overestimate the energy of the system. Moreover, the
minima of the variational energy will not lie exactly at the same
positions in parameter space as those of the true energy. As we
increase the number of modes, we will get closer to the true result,
and if our function system were complete, we could be certain of full
convergence of the variational results to the correct answer. We have
no formal proof of the completeness of the system of multi-cycloids
but note that as a function of $\xi$, the systems used for the
representation of the abscissa and the ordinate of the curve are
complete in the spaces of odd and even functions, respectively.
Completeness is difficult to prove because of the correlation between
the coefficients describing the abscissa and the ordinate. However, we
suspect that for all practical purposes of representing curves that
resemble a cycloid as closely as do the numerically obtained
solutions, our function system can be considered complete.

In the full numerical computation \cite{spencer:94}, with which we
compared our results, the discretization of abscissae is given by a
formula akin to \eqref{eq:x} with an equidistant distribution of the
parameter $\xi$ and the interface position is given as a superposition
of cosine modes in the same parameter. Hence, numerical convergence
relies on the assumption of completeness of a function system derived
from Fourier modes by a stretching in the $x$ coordinate.

As long as the higher modes have small enough amplitudes, the two
approaches should give equivalent results. Since we solve the elastic
problem essentially analytically and for a continuous interface, not a
discrete one, we reach the same accuracy as the numerics with fewer
modes.

Note that the cusp singularity is {\em not} an inherent restriction to
the method, as the function system is chosen such that it can
represent one (or several) cusps. However, when a cusp appears,
quantities such as the elastic energy density diverge there. This
means that the numerical solution of the nonlinear system of equations
\eqref{eq:dritte} for the variational parameters will run into
problems, hence the cusp cannot be reached exactly in this final
numerical step.

Even the one-mode approximation suggests the existence of stable
large-amplitude steady states in the presence of gravity, as is
demonstrated by Fig.~\ref{fig:monozyk3}. Taking into account more
modes, we obtain a quantitatively satisfactory description of the
numerical Spencer-Meiron branch, conveying some confidence that the new
branches with gravity are equally well described by this approach. The
stability domain suggested by the one-mode picture is roughly confirmed
in the three-mode representation (see Fig.~\ref{fig:multizyk}).  As
gravity is increased, there are no small-amplitu\-de stable solutions
anymore.  At first sight, this might seem counterintuitive: why should
gravity, a stabilizing effect, destroy the stability of small-amplitude
solutions? The answer is that gravity renders the zero-amplitude, i.e.
planar, solution more stable and hence larger amplitudes are needed for
true structures to become stable.

Hence, we conclude that in {\em confined} systems under gravity or a
similar body force (it has been shown that in directional
solidification a temperature gradient acts just as a strong effective
gravity field \cite{durand:96,cantat:98}) stable steady states may
exist at large amplitude and be absent at small ones.

Below the instability threshold, i.e., for parameters where the planar
interface is stable, the system may be forced into a large amplitude
state by a sufficiently strong perturbation. This clearly calls for
numerical simulations and experimental attempts at creating these
states.

In {\em extended} systems, the absence of stable steady-state
solutions at large wavelength as well as the numerical evidence from
time-dependent simulations \cite{spencer:94,kassner:01} suggest that
the cusp singularity is indeed reached in finite time. This is of
course a statement within linear elasticity theory.  It means that
stresses would increase beyond all limits in the minimimum of a
groove, if linear elasticity held all the time. If linear elasticity
were valid up to the fracture threshold, one might conclude from this
result that the Grinfeld instability would inevitably lead to fracture
in such a situation. However, the answer to this question is beyond
the scope of this paper, as it is obvious that at sufficiently large
stresses plasticity must be taken into account. Phase-field
simulations containing an inherent yield stress in the model
\cite{kassner:01} suggest that indeed cracking is a likely scenario in
sufficiently extended systems.

\begin{acknowledgement}
  This work was supported by the \textit{Deutsche
    Forschungsgemeinschaft} under Grant No.~Ka 672/4-2 and FOR
  301/2-1, which is gratefully acknowledged. In addition, we
  acknowlede travel grants by PROCOPE, Grant No.~9619897 (DAAD,
  Germany) and 97176 (APAPE, France), enabling a closer collaboration
  between the two groups involved in this work.
\end{acknowledgement}

\appendix

\section{\label{app:generalmapping}
Calculation of the strain energy density along a multi-cycloid surface}

Instead of calculating the strain energy density inside the bulk as in
\cite{kohlert:02a}, we will only describe here how to calculate this
energy density $w(s)$ at the surface. We carry out the calculation for
general $N$-cycloids.

In general, the elastic energy density of a solid submitted to plane
strain can be written in terms of the two-dimensional stress tensor as
\begin{equation}
w=\frac{1+\nu}{2E} \left(\sigma_{ij} \sigma_{ij} - \nu\, \sigma_{kk}^2\right)\>,
\end{equation}
where summation over repeated subscripts is implied. At the interface,
$\sigma_{ij}$ is diagonal with elements $\sigma_{tt}$ and
$\sigma_{nn}$. Since $\sigma_{tt}={\rm Tr\,} \sigma -\sigma_{nn}$ and
because in our normalization $\sigma_{nn}=0$, the elastic energy
density can be expressed by the trace of the stress tensor alone,
which allows us to deal with a single scalar. Thus the strain energy
density takes the form
\begin{equation}
w(x,y)=\frac{1}{2}\left({\rm Tr\,}\sigma\right)^2\>,
\end{equation}
where ${\rm Tr\,}\sigma$ is in our non-dimensional scalings equal to
$1\!+\!\sigma_{xx}\!+\!\sigma_{yy}$.

The idea is to employ a mapping of the half-plane bounded from above
by our multi-cycloid onto the area below the real axis using the
analytic function
\begin{equation}
z=\omega(\varsigma)=\varsigma-i\sum_{n=1}^{N}\frac{\rho_n}{nk} e^{-ink\varsigma}
\end{equation}
where $\varsigma=\xi+i\eta$. This defines a mapping of the domain
$\Im(z)\le \zeta$ to the lower $\varsigma$ half plane, as can be seen
easily by restricting to $\varsigma=\xi$ which shows that the
interface is mapped to the real axis. In order to solve the elastic
problem we have to satisfy \cite{kassner:01}
\begin{equation}
\label{eq:thecrucialone} \phi_0(\xi)+\left[\omega(\xi)
-\overline{\omega}(\xi)\right]\frac{\overline{\phi'_0}(\xi)}
{\overline{\omega'}(\xi)}+\overline{\psi'_0}(\xi)
=\frac{\overline{\omega}(\xi)-\omega(\xi)}{2} \>,
\end{equation}
where $\phi_0$ and $\psi_0$ are modified Goursat functions. These
functions must be analytic functions of $\zeta$ for $\eta\to-\infty$,
so it has to be established that ${\psi'_0}(\xi)$ contains no
exponentials increasing for $\eta\to -\infty$ when $\xi$ is replaced
by $\xi+i\eta$.  Since $\overline{\psi'_0}(\xi)$ is the complex
conjugate of a function that is analytic in the {\em lower} half
plane, it must be analytic in the {\em upper} half plane, which means
that terms of the form $\exp(in\xi)$ are allowed whereas
$\exp(-in\xi)$ are not (for details see \cite{kassner:01}). For
brevity, we will designate the forbidden terms as ``negative
exponentials''. Technically, we make an ansatz for $\phi_0(\xi)$:
\begin{equation}
\phi_0(\xi)=\frac{i}{k}\sum_{n=1}^{N}\alpha_{n}e^{-ink\xi}\>,
\end{equation}
where we may assume $\alpha_n$ to be real (which will be justified
later). Now let us simplify Eq.~\eqref{eq:thecrucialone}. We have
\begin{equation}
\left[\omega(\xi)-\overline{\omega}(\xi)\right] =2i\Im\left(\omega(\xi)\right)
=-\frac{2i}{k}\sum_{n=1}^{N}\frac{\rho_n}{n}C_n
\end{equation}
and hence
\begin{equation}
\label{eq:umgestellt}\overline{\psi'_0}(\xi)
=\frac{1}{k}\sum_{n=1}^{N}\frac{\rho_n}{n}C_n
 \left(1+2\frac{\overline{\phi'_0}(\xi)}
{\overline{\omega'}(\xi)}\right) -\phi_0(\xi) \>.
\end{equation}
Via the choice of $\alpha_n$ we have to establish that the right hand
side of Eq.~\eqref{eq:umgestellt} contains no negative exponentials.
Let us further simplify the representation. We have
\begin{subequations}
\begin{equation}
\overline{\phi'_0}(\xi)=\sum_{n=1}^{N}n\alpha_n e^{ink\xi}\>,
\end{equation}
\begin{equation}
\overline{\omega'}(\xi)=1-\sum_{n=1}^{N}\rho_n e^{ink\xi}\>.
\end{equation}
\end{subequations}
For the division, we use the common expression for the quotient of two
series (\cite{abra:1},\,p.\,28): Let
\begin{eqnarray*}
s_1&=&1+a_1x+a_2x^2+a_3x^3+\dots \>, \\
s_2&=&1+b_1x+b_2x^2+b_3x^3+\dots \>, \\
s_3&=&1+c_1x+c_2x^2+c_3x^3+\dots \>, \\
s_4&=&1+d_1x+d_2x^2+d_3x^3+\dots \>, \\
s_5&=&1+\epsilon_1x+\epsilon_2x^2+\epsilon_3x^3+\dots \>,
\end{eqnarray*}
with \[s_1=1+\overline{\phi'_0}(\xi), \quad
s_2=\overline{\omega'}(\xi)\>,\] that is
\[a_n=n\alpha_n,\quad b_n=-\rho_n,\quad x=e^{ik\xi}\>.\]
We then need to calculate
\begin{eqnarray*}
s_3&=&\frac{s_1}{s_2} \>, \\
s_4&=&{s_2}^{-1} \>, \\
s_5&=&s_3-s_4 \>=\> \frac{\overline{\phi'_0}}{\overline{\omega'}} \>.
\end{eqnarray*}
The coefficients $c_n$ and $d_n$ are given by the recursion
\begin{eqnarray*}
c_1&=&a_1-b_1\>,\\
c_n&=&a_n-\left[b_n+\sum_{j=1}^{n-1}b_jc_{n-j}\right]\>,\\
d_1&=&-b_1\>,\\
d_n&=&-\left[b_n+\sum_{j=1}^{n-1}b_jd_{n-j}\right]\>.
\end{eqnarray*}
This leads to
\begin{equation}
\begin{split}
\epsilon_1=& a_1\>,\\
\epsilon_n=& a_n-\sum_{j=1}^{n-1}b_j\epsilon_{n-j}\>.
\end{split}
\end{equation}
We then have, after introducing $\epsilon_0=\nicefrac{1}{2}$,
\begin{equation}
1+2\frac{\overline{\phi'_0}(\xi)} {\overline{\omega'}(\xi)}
 =1+2s_5 =2\sum_{n=0}^\infty
\epsilon_nx^n\>.
\end{equation}
Therefore,
\begin{equation}
\begin{split}
\epsilon_0=&\frac{1}{2}\>,\qquad\epsilon_1=\alpha_1 \>,\\
\epsilon_n=& n\alpha_n+\sum_{j=1}^{n-1}\rho_{j}\epsilon_{n-j}\>,
 \quad n=2\dots N\>.
\end{split}
\end{equation}
Negative exponentials on the right-hand side of \eqref{eq:umgestellt}
can only result from the negative exponentials in
\[C_n=\frac{1}{2}\left(e^{ink\xi}+e^{-ink\xi}\right)\>.\]
We cut out the relevant parts from Eq.~\eqref{eq:umgestellt},
consisting of negative exponentials and require them to become zero,
to compute the coefficients determining $\phi_0$ in terms of the
$\rho_n$
\begin{equation}
\left(\sum_{m=1}^{N}\frac{\rho_m}{m}e^{-imk\xi}\right)\left(\sum_{j=0}^{N-1}
\epsilon_je^{ijk\xi}\right)=\sum_{n=1}^{N}\alpha_{n}e^{-ink\xi}\>,
\end{equation}
with $\epsilon_0=\nicefrac{1}{2}$ and the prefactor $ik^{-1}$ dropped.
Now we sort the terms in this equation by exponentials which finally
gives us the system
\begin{equation}
\alpha_n=\sum_{j=0}^{N-n}\epsilon_j\frac{\rho_{n+j}}{n+j}\>,\quad n=1\dots N \>.
\end{equation}
Note that the $\epsilon_j$ contain the $\alpha_i$ as well, so the
equations are not a simple recursive scheme but a linear system of
equations for the $\alpha_n$. Now we get back to ${\rm Tr\,}\sigma$,
which can be written as
\begin{equation}
{\rm Tr\,}\sigma=1+4\frac{\dot{x}(\xi)\Re(\phi'(\xi))+ \dot{y}(\xi)\Im(\phi'(\xi))}
{\dot{x}(\xi)^2+\dot{y}(\xi)^2} \>.
\end{equation}
The denominator has already been written down during the calculation
of the arc length.  The numerator can be simplified in a similar
manner. After all simplifications have been performed, we finally get
\begin{equation}
\begin{split}
\label{eq:diewirklicheenergiedichte}
 w(s) &= \frac{1}{2}\left(1+4\frac{w_1(s)}{1+w_2(s)}\right)^2\\
 w_1(s) &= \sum\limits_{n=1}^N n\alpha_n\left(C_n-\rho_n\right)\\
  &\mbox{}-\sum\limits_{n=1}^{N-1}C_n
  \sum\limits_{j=1}^{N-n}\left(\rho_{j}\alpha_{j+n}(j+n)
  +\rho_{j+n}\alpha_{j}j \right)\\
 w_2(s) &= \sum\limits_{n=1}^{N}\rho_n(\rho_n-2C_n)
  +2\sum\limits_{n=1}^{N-1}C_n\sum\limits _{j=1}^{N-n}\rho_{{j}}
  \rho_{{j+n}}\>.
\end{split}
\end{equation}

\section{\label{app:howtosolve}Details of the numerical solution of
Eq.~(\ref{eq:dritte})}

Let us first repeat the system of partial differential equations
\eqref{eq:dritte}.

\begin{equation}
\label{app:problem} \frac{\partial}{\partial\rho_j}\left[E_{\rm e}+E_{\rm s}+E_{\rm
g}\right]=0,\quad j=1\dots N\>.
\end{equation}
Herein, the energy changes entering the equation are integrals over
certain rational terms containing trigonometric functions and the
vector of amplitudes $\underline{\rho}=(\rho_1\dots\rho_N)$ as well as
the wave number $k$ and the gravity parameter $l_{12}$.

We first reformulate the problem \eqref{app:problem} by considering
the fact that $k$ is contained in the energy terms as a prefactor only
(compare Eqs.~\eqref{eq:dEen}, \eqref{eq:dEgn} and
\eqref{eq:dEsn})\footnote{De facto we have also transformed the
  integration variable $\xi\to X=k\xi$, changing the integration
  interval to $[0,\pi]$.}.  Then we get a simplified problem with
modified $E$ terms:
\begin{equation}
\label{app:problem2}\frac{\partial}{\partial \rho_j}\left[\frac{1}{k}
 \tilde{E}_{\rm e}(\underline{\rho})
 + \tilde{E}_{\rm s}(\underline{\rho})
 + \frac{1}{k^2}\tilde{E}_{\rm g}(\underline{\rho})\right]=0 \>.
\end{equation}

The question is now to specify which subset of the solution manifold
is required. We concentrate on fixed physical system parameters, i.e.,
constant $l_{12}$ in order to produce the lines shown in
figure~\ref{fig:multizyk}.

Next a suitable numerical method has to be chosen. Solutions are known
to satisfy $\underline{\rho}\approx\underline{0}$ for
$\overline{\alpha}\approx 0$, and solutions starting a branch are
given by linear stability analysis:
$k_{\mathrm{start}}=1+\sqrt{1-l_{12}}$; $\rho_{j,\,\mathrm{start}}=0$,
$j=1\dots N$.

As solutions along a branch are expected to change continuously, we
can implement a Newton Raphson algorithm and move along the selected
branch by varying a parameter. Yet, already in the monocycloid model
we find that some of the curves are multi-valued with respect to $k$.
This renders it unfavourable to use $k$ as a fixed parameter in that
scheme and to solve for $\rho_1$, because the exact turning points are
unknowns.

It turns out that for all branches situated between $k=1$ and $k=2$,
and with $l_{12}\in [0,1]$, the curves behave monotonously as a
function of $\rho_1$ up to the cusp. These are the cases exhibited in
figure~\ref{fig:multizyk}. Therefore, in order to solve the system of
equations \eqref{app:problem2} for $j=1\dots N$, we keep $\rho_1$
fixed instead of $k$ and take $\{k, \rho_2\dots\rho_N\}$ as our set of
variables to be determined by the iteration. This results in a
modified Jacobian containing terms
$\{\partial^2{E}/\partial\rho_n\partial{k}$,
$\partial^2{E}/\partial\rho_n\partial\rho_m\}$ $(m=2\dots N,n=1\dots
N)$.

As initial guess for the first non-zero solution of a branch, we
choose the upper marginal wavenumber from linear theory for $k$, a
small value of $\rho_1$ and set all other $\rho_i$ equal to zero.
After having found the first solution, we move along the solution
branch towards larger $\rho_1$ values in steps of typically
$\Delta\rho_1=0.001$.  Consecutive data sets are estimated by forward
differences using up to six solution points and then iterated until
the 2-norm of the vector of the changes remains below a threshold of
typically $10^{-10}$.

It should be emphasized that the derivatives of Eqs.~\eqref{eq:dEen},
\eqref{eq:dEgn} and \eqref{eq:dEsn} can be given analytically (they
are omitted here because they are rather lengthy expressions), and so
the Newton Raphson algorithm can be programmed with an (up to
quadrature) exact Jacobi matrix. This makes the code converge
extremely fast.

A more thorough investigation of the solution manifold with respect to
small wavenumbers and $l_{12}>1$ goes beyond the scope of this paper
because monotony considerations do not apply in this range and
multi-valuedness of the manifold can appear which is connected to
coarsened solutions. These considerations will be presented in a
separate work.

\bibliographystyle{unsrt}

\end{document}